\documentclass[preprint,12pt]{elsarticle}

\usepackage{amssymb}
\usepackage{amsthm}
\usepackage{amsmath}
\usepackage{algorithm}
\usepackage{fourier}
\usepackage{algpseudocode}
\biboptions{numbers,sort&compress}
\journal{Computer Aided Geometric Design}

\begin{document}

\begin{frontmatter}



\title{Reliability-based $G^{1}$ Continuous Arc Spline Approximation}
\author[1]{Jinhwan Jeon}
\ead{jordan98@kaist.ac.kr}
\author[1,2]{Yoonjin Hwang}
\ead{yoonjinh@kaist.ac.kr, yoonjin.hwang@hankookn.com}
\author[1]{Seibum B. Choi\corref{cor1}}
\ead{sbchoi@kaist.ac.kr}

\cortext[cor1]{Corresponding Author}
\affiliation[1]{organization={Department of Mechanical Engineering, KAIST},
            addressline={291 Daehak-ro Yuseong-gu},
            city={Daejeon},
            postcode={34141},
            country={Republic of Korea}}

\affiliation[2]{organization={Hankook \& Company},
            addressline={286 Pangyo-ro Bundang-gu},
            city={Seongnam},
            postcode={13494},
            country={Republic of Korea}}

% \author{}

% \affiliation{organization={},%Department and Organization

\begin{abstract}
In this paper, we present an algorithm to approximate a set of data points with $G^{1}$ continuous arcs, using points' covariance data. To the best of our knowledge, previous arc spline approximation approaches assumed that all data points contribute equally (i.e. have the same weights) during the approximation process. However, this assumption may cause serious instability in the algorithm, if the collected data contains outliers. To resolve this issue, a robust method for arc spline approximation is suggested in this work, assuming that the 2D covariance for each data point is given. Starting with the definition of models and parameters for \textbf{single arc approximation}, the framework is extended to \textbf{multiple-arc approximation} for general usage. Then the proposed algorithm is verified using generated noisy data and real-world collected data via vehicle experiment in Sejong City, South Korea.
\end{abstract}

\begin{graphicalabstract}
    \centering
    \includegraphics[width=0.8\columnwidth]{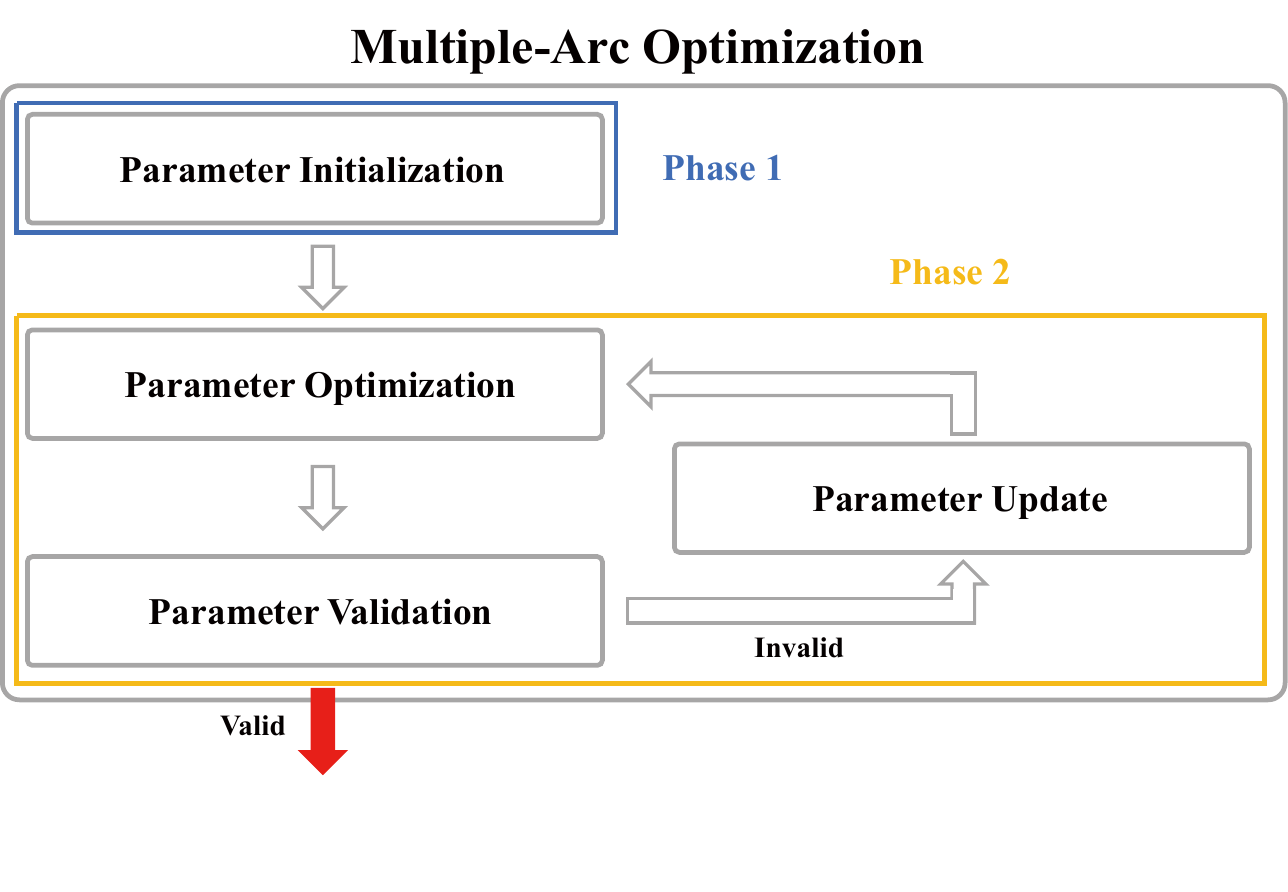}
    \includegraphics[width=0.9\columnwidth]{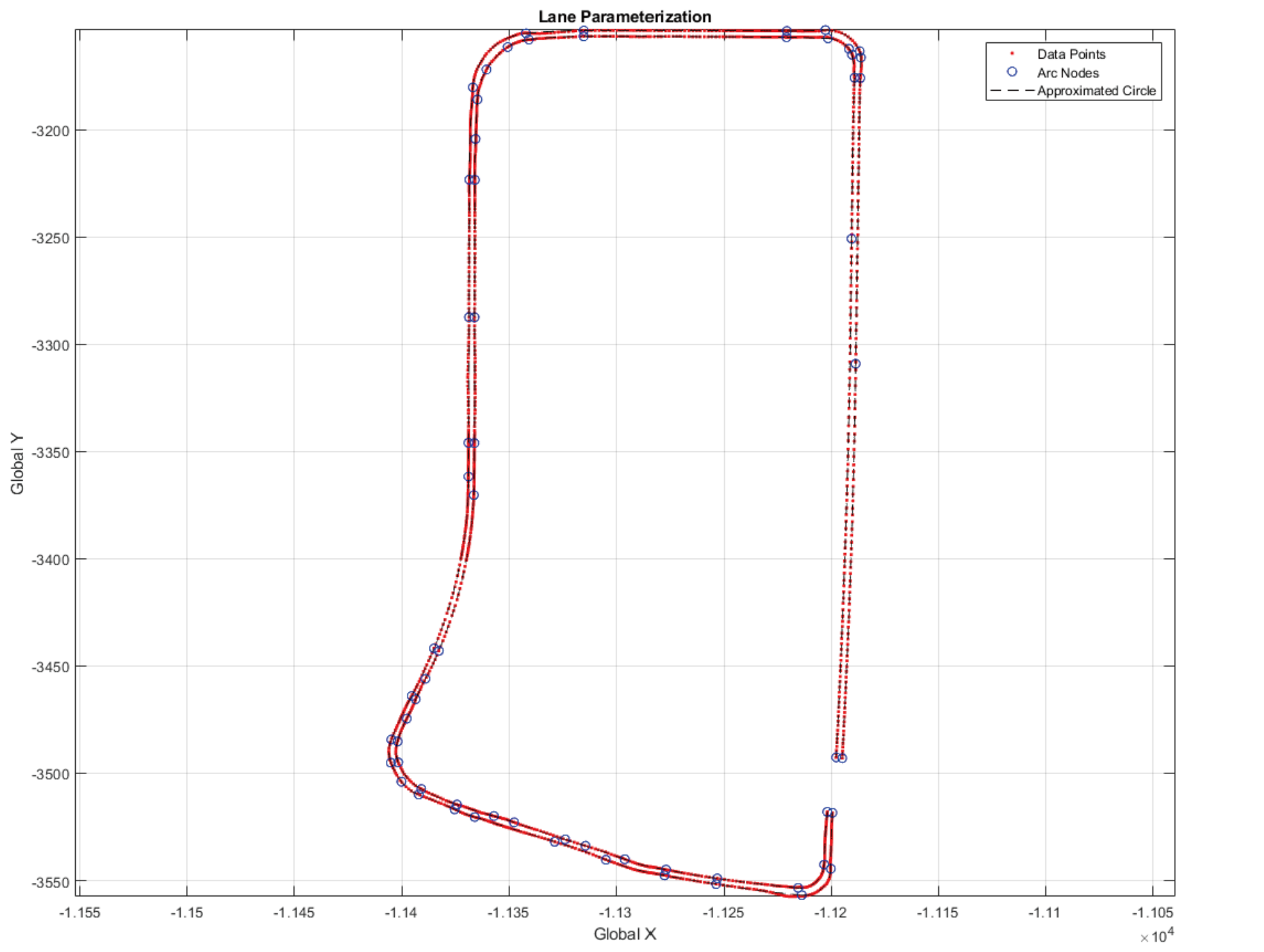}
\end{graphicalabstract}

\begin{highlights}
\item Proposal of model for single arc optimization
\item Proposal of an optimization framework for multiple arc optimization
\item Verification of proposed optimization algorithm via real-world dataset
\end{highlights}
% \begin{highlights}
% \item Proposal of model for single arc optimization
% \item Proposal of an optimization framework for multiple arc optimization
% \item Verification of proposed optimization algorithm via real-world dataset
% \end{highlights}


\end{frontmatter}

% \linenumbers

\section{Introduction}
\label{Chap1}
Various efforts have been made to analyze the tendency in point data or to smoothen a sequence of points using different families of curves. Representative fields of applications are numerically controlled (NC) machining\cite{MEEK1995,Jee2003,Kim2007}, curve reconstruction\cite{LEE2000}, and road lane data parameterization\cite{Schindler2011,Schindler2012,MAIER2014}. Many different types of curves were presented for these applications\cite{Klass1983,PIEGL1987,HOSCHEK1992}. Among these curves, making use of arcs for data approximation is especially more attractive due to its simplicity and translation/rotation invariant properties.

\subsection{Literature Review and Problem Statement}
A lot of meaningful research was done on arc spline approximation using biarcs\cite{MEEK1992,YANG1996,PIEGL2002,Heimlich2008,Song2009}. For example, in the work of \cite{Song2009}, a single 3D arc was identified with a position vector $(\mathbb{R}^{3})$, 2 length-related parameters $(\mathbb{R})$, and ZXZ-Euler angles to form rotational matrix $\in SO(3)$. With these parameters, a single arc can be optimized without any constraints, i.e., unconstrained optimization. Unconstrained optimization has a huge advantage over constrained optimization-based methods since unconstrained optimization generally shows more accurate and faster convergence. Using biarc interpolation between two arcs, \cite{Song2009} succeeded in parameterizing data points into multiple arc segments, without performing constrained optimization. However, there were some drawbacks to this method. When given data points are very close to a line, the length parameter becomes infinitesimally small, causing near-singularity problems during Gaussian step computation. Moreover, in order to perform optimization using biarcs, about $\frac{1}{4} n$ arc segments were needed for parameterization, where $n$ is the number of data points. This implies that, on average, one arc segment is allocated for 4 data points, which is not a remarkable result, considering the original purpose of parameterization(i.e. compact data representation). 

Other than compact data representation, previous research also suffers from algorithm instability due to the inherent noise in the data collection process. RANSAC\cite{Fischler1981} is typically used when removing outlier or noisy points from a dataset\cite{Schindler2012}, but there are two main critical drawbacks: there is no upper bound on the time it takes to perform robust regression, and there is no guarantee that all the outlier points are removed after RANSAC. Other than using RANSAC, in the work of \cite{Drysdale2008,Heimlich2008,MAIER2014}, they defined tolerance channels as a lateral offset of data points. In these researches, the collected data points are assumed to be highly accurate(low level of noise), and the arc spline approximation is performed so that approximating arcs do not exceed the bounds defined by the tolerance channels. Since tolerance channels are generated by adding equal lateral offsets to all of the data points, noisy data will cause a significant decrease in algorithm performance and stability.

\subsection{Overview of Our Approach}
We aim to solve arc spline approximation problems with noisy data both compactly and robustly. In order to work around the limitations of previous works, we focused on the concept that each data point should contribute with different importance when performing arc spline approximation. Assuming that 2D covariance is given for each data point, we have constructed an optimization problem that reflects this idea. 

The paper is organized as follows: In \textbf{section \ref{Chap2}}, the method of approximating data points with a single arc is explained. Specifically, parameters that form an arc will be introduced, and cost function models for single arc approximation are proposed. Then, the proposed single arc approximation will be tested with both generated and real-world dataset. In \textbf{section \ref{Chap3}}, models discussed in section \ref{Chap2} are extended for the implementation of the multiple-arc approximation framework. Since we are formulating data approximation (fitting) as a nonlinear optimization problem, optimization variables should be initialized appropriately for stable convergence. Moreover, after finding the optimal arc parameter set for a fixed number of arc segments, we increase the number of arcs if the current number of arc segments is not enough for accurate data approximation. The variable initialization and arc parameter validation/update procedure mentioned above will be explained thoroughly in section \ref{Chap3}. Similar to section \ref{Chap2}, evaluation of the proposed multiple arc approximation framework will be done with real-world collected data points and covariance. Finally in \textbf{section \ref{Chap4}}, possible applications of the proposed optimization framework and future research directions are addressed. 

\subsection{Contributions}
Our main contributions are as follows:
\begin{itemize}
    \item Cost function/constraint models for \textbf{Single Arc Approximation}
    \item Arc parameter initialization for \textbf{Multiple Arc Approximation}
    \item Cost function/constraint models for \textbf{Multiple Arc Approximation}
    \item Arc parameter validation/update for \textbf{Multiple Arc Approximation}
\end{itemize}

\newpage
\section{Single Arc Approximation}
\label{Chap2}
Before jumping into general multiple-arc approximation, an algorithm for single-arc approximation will be discussed in this chapter. In detail, parameters that form an arc will be defined, and a novel cost function/constraint model will be explained. Note that the parameter and cost functions defined in this section will be used for multiple-arc approximation, with a slight modification.

\subsection{Parameters for Single Arc}

\begin{figure}[t]
    \centering
    \includegraphics[width=0.55\columnwidth]{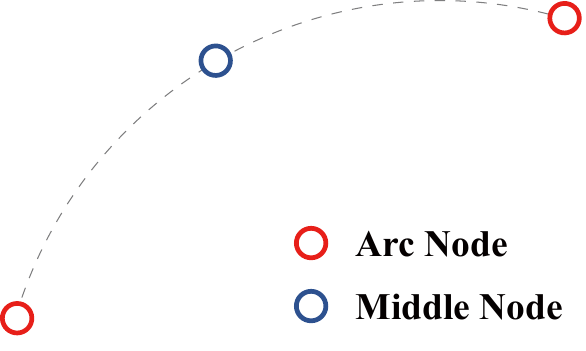}
    \caption{3 Points are set as parameters(optimization variables) to define a \textbf{single arc}}
    \label{fig:SingleArc}
\end{figure}

 There are various parameter combinations to represent a single arc\cite{HOSCHEK1992, Drysdale2008, Song2009, MAIER2014}. Since we are tackling data approximation via the construction of a nonlinear optimization problem, optimization variables should be similar in numerical scale for stable and accurate convergence. Extreme discrepancies in the scales of variables, where one becomes extremely small and the other very large, can result in significant numerical errors during arithmetic operations. Such errors may lead to algorithm divergence or a degradation in accuracy. Considering this condition, defining an arc with three points and setting these points as the optimization variables would be the best option since the positional vectors of three points have similar numerical scales. Among these three points, the first and the second points belong to the start and the end points of the arc respectively, and the third point defines the middle of the arc segment. As shown in figure \ref{fig:SingleArc}, the two red points serve as \textbf{arc nodes}, and the blue point at the middle of the arc segment is referred to as the \textbf{middle node} throughout the paper. The positions of these three points are designated as optimization variables, and the optimization process aims at refining their position to achieve the best-fit single arc for the given data points and covariance.

\subsection{Optimization Models}
In order for arc parameters (2 Arc nodes and 1 Middle node in figure \ref{fig:SingleArc}) to approximate data points accurately, appropriate modeling of cost function is essential. A total of 3 models are augmented to form the full cost function for \textbf{single arc approximation}, and the final goal is to find the optimal arc parameters that minimize the augmented cost function. 

\begin{figure}[t]
    \centering
    \includegraphics[width=0.65\columnwidth]{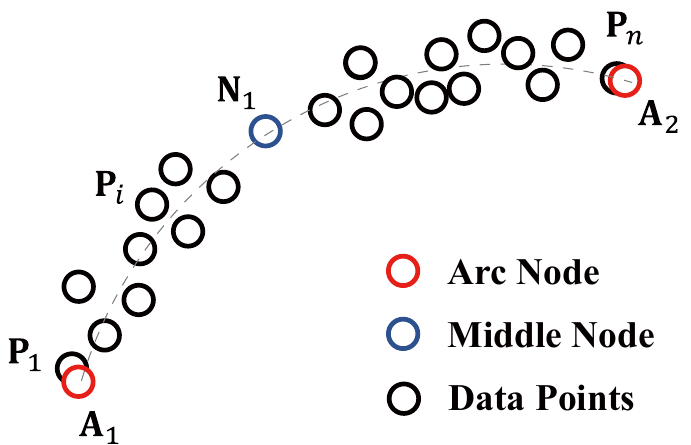}
    \caption{Anchor Model for Single Arc Approximation: Arc Nodes ($\mathbf{A}_{1}, \mathbf{A}_{2}$) are matched with first and last data points ($\mathbf{P}_{1}, \mathbf{P}_{n}$) respectively. Middle node is not included in the anchor model.}
    \label{fig:SingleArcAC}
\end{figure}

\subsubsection{Single Arc Anchor Model}
The first cost function model is called the anchor model. The main purpose of this part of the cost function is to "anchor" the arc nodes to the first and the last data points for stable optimization, as shown in figure \ref{fig:SingleArcAC}. Assuming that the data points are in order, the first and second arc nodes are matched to the first and the last data points respectively. To match each arc node with the corresponding data point, the anchor model cost is defined as the sum of squared, weighted Euclidean distance between the matched points. The anchor model cost is written as follows.

\begin{equation}
\begin{aligned}
    \mathcal{L}_{\mathrm{AC}} = & {\lVert \mathbf{P}_{1} - \mathbf{A}_{1} \rVert}^{2}_{\Sigma_{\mathrm{AC}}} + {\lVert \mathbf{P}_{n} - \mathbf{A}_{2} \rVert}^{2}_{\Sigma_{\mathrm{AC}}} \\
    = & {\left( \mathbf{P}_{1} - \mathbf{A}_{1} \right)}^{\top}{\Sigma_{\mathrm{AC}}}^{-1} {\left( \mathbf{P}_{1} - \mathbf{A}_{1} \right)} + {\left( \mathbf{P}_{n} - \mathbf{A}_{2} \right)}^{\top}{\Sigma_{\mathrm{AC}}}^{-1} {\left( \mathbf{P}_{n} - \mathbf{A}_{2} \right)}
\end{aligned}
\label{eq:single_AC}
\end{equation}

\noindent
$\mathbf{P}_{1}$ and $\mathbf{P}_{n}$ are the first and the last points in the dataset, $\mathbf{A}_{1}$ and $\mathbf{A}_{2}$ stand for the first and second arc nodes respectively. The notation ${\lVert \cdot \rVert}^2_{\Sigma} = (\cdot)^{\top} \Sigma^{-1} (\cdot)$ denotes the squared Mahalanobis Distance, which can be thought of as the square of weighted Euclidean distance. The weight is reflected in the cost by additionally multiplying the inverse of the covariance matrix $\Sigma_{\mathrm{AC}}$ to the squared 2-norm of the original residual vectors $\mathbf{P}_{1} - \mathbf{A}_{1}$ and $\mathbf{P}_{2} - \mathbf{A}_{2}$, as shown in equation \ref{eq:single_AC}. 

\begin{figure}[t]
    \centering
    \includegraphics[width=0.52\columnwidth]{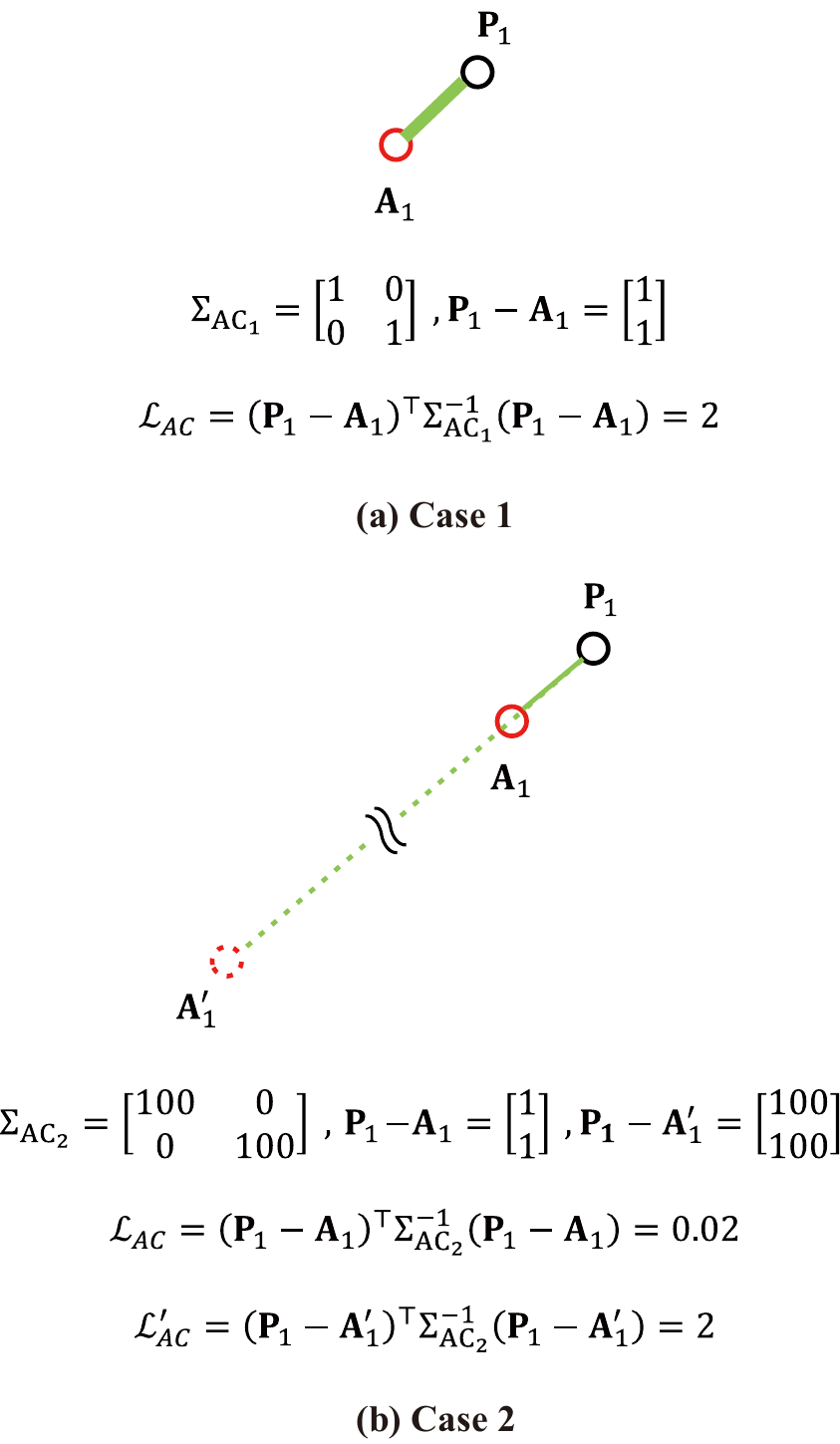}
    \caption{Anchor Model for Cases: (a)$\Sigma_{\mathrm{AC}_1}=\left[\begin{array}{ll}1 & 0 \\ 0 & 1\end{array}\right]$ (b)$\Sigma_{\mathrm{AC}_2}=\left[\begin{array}{cc}100 & 0 \\ 0 & 100\end{array}\right]$}
    \label{fig:SingleArcAC1}
\end{figure}

\hfill \break
\noindent \textbf{Remarks on Anchor Model Covariance $\Sigma_{\mathrm{AC}}$}

In addition, the anchor model cost can be controlled by tuning the covariance matrix $\Sigma_{\mathrm{AC}}$ in equation \ref{eq:single_AC}. For example, let us set $\Sigma_{\mathrm{AC_{1}}}$ as an identity matrix initially, as shown in figure \ref{fig:SingleArcAC1}(a): Case 1. If one increases the diagonal terms of $\Sigma_{\mathrm{AC_{1}}}$ to 100 (figure \ref{fig:SingleArcAC1}(b): Case 2), this will lead to cost reduction by $\frac{1}{100}$ of the original anchor model cost. This is due to the fact that we multiply the inverse of the covariance matrix when determining the weighted cost in equation \ref{eq:single_AC}. We can also observe in figure \ref{fig:SingleArcAC1}(b): Case 2 that, point $\mathbf{A}_{1}$ is shifted to $\mathbf{A}^{'}_{1}$ so that residual vector $\mathbf{P}_{1} - \mathbf{A}^{'}_{1}$ is set to be 100 times larger than the residual of Case 1. However, due to the covariance difference of the two cases, the resultant cost is the same. Thus, we can imply that increasing the diagonal terms of $\Sigma_{\mathrm{AC}}$ "permits" the arc nodes to be located further away from the matched data points without increasing the anchor model cost. This ultimately means that the anchor model becomes less sensitive to (arc node, data point) paired matching error.

Finally, it is important to highlight that the middle node (Blue node marked $\mathbf{N}_{1}$) in figure \ref{fig:SingleArcAC} is not included in the anchor model cost. The reason for omission is that the middle node needs to move freely around in space during optimization iterations. Given its significant role in determining the arc's radius, anchoring it to a specific point in the dataset is considered to be inappropriate.  

\subsubsection{Single Arc Measurement Model}
\label{Chap2ME}
The second cost function model is called the arc measurement model. Its goal is to minimize the error between data points and the approximating arc. Unlike the previously mentioned anchor model, the arc measurement model incorporates the middle node position in its cost calculation. In the overall cost function, where the anchor model is responsible for anchoring the two arc nodes, the primary focus of the arc measurement model is to adjust the middle node's position, determining the optimal "shape" of the arc.

\begin{figure}[t]
    \centering
    \includegraphics[width=0.8\columnwidth]{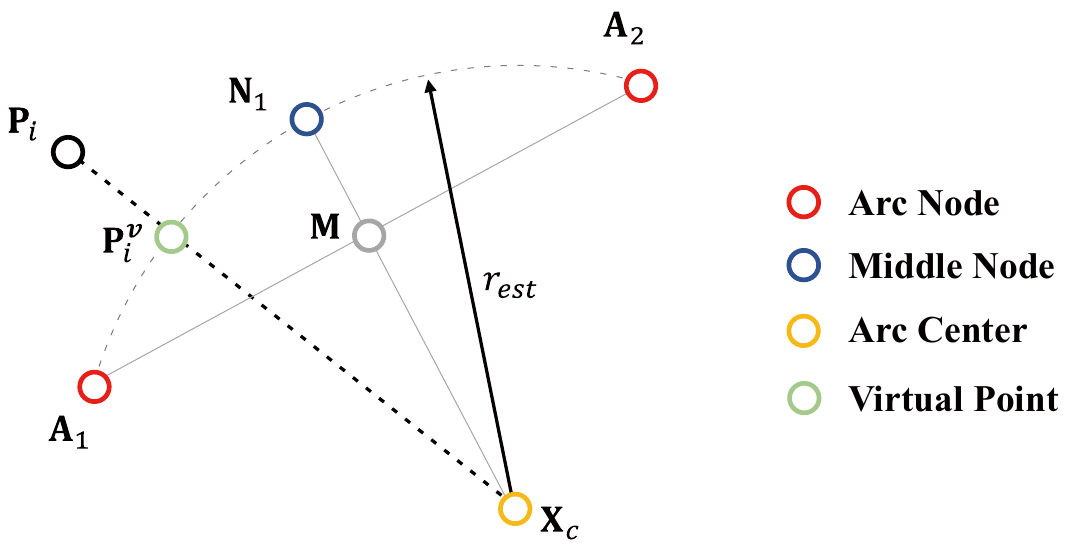}
    \caption{Arc Measurement Model for Single Arc Approximation: (1) Arc center $\mathbf{X}_{C}$ is computed geometrically, by using two Arc Nodes and the Middle Node. (2) A virtual point(marked green) $\mathbf{P}_i^{v}$ is defined by finding the intersection of line $\mathbf{P}_{i}\mathbf{X}_{C}$ and arc $\widearc{A_{1}N_{1}A_{2}}$. (3) Matching error (model residual) $\mathbf{r}^{i}_{\mathrm{ME}}$ is computed by subtracting data point vector $\mathbf{P}_{i}$ from virtual point vector $\mathbf{P}^{v}_{i}$. (4) Residual is weighted using the pre-obtained covariance matrix $\Sigma^{i}_{\mathrm{ME}}$ of point $\mathbf{P}_{i}$. (5) Arc measurement model cost is accumulated as $i$ iterates from 1 to $n$, the data size.}
    \label{fig:SingleArcME}
\end{figure}

\hfill \break
\noindent \textbf{Cost Computation}

The process of computing the arc measurement model cost is as follows. Refer to figure \ref{fig:SingleArcME} for graphical understanding.
\begin{enumerate}
    \item Compute the position of arc center $\mathbf{X_{C}}$ from the positions of two Arc Nodes and the Middle Node, using simple geometry.
    \item Set loop variable $i$ to iterate from 1 to $n$ (data size). For example, point $\mathbf{P}_{i}$ is chosen for explanation.
    \item Find the intersection of line $\mathbf{P}_{i}\mathbf{X}_{C}$ and arc $\widearc{A_{1}N_{1}A_{2}}$, and set this point as the virtual point $\mathbf{P}^{v}_{i}$.
    \item Arc measurement model residual $\mathbf{r}^{i}_{\mathrm{ME}}$ is defined as the difference between data point $\mathbf{P}_i$ and virtual point $\mathbf{P}_i^{v}$ : $\mathbf{r}^{i}_{\mathrm{ME}} = \mathbf{P}_i^{v} - \mathbf{P}_i$.
    \item Using the covariance $\Sigma^{i}_{\mathrm{ME}}$ of point $\mathbf{P}_{i}$ obtained beforehand, residual $\mathbf{r}^{i}_{\mathrm{ME}}$ is weighted (squared Mahalanobis distance). 
    \item Steps 3 to 5 are repeated for the whole dataset($i$ iterating from 1 to $n$). The arc measurement model sums up all the costs computed in step 5. 
\end{enumerate}

Before having an in-depth explanation of the cost computation process, note that we assume 2D covariance data for each data point is already computed beforehand. The detailed process of obtaining covariance data for each data point will be explained in chapter \ref{Chap4}. 

In figure \ref{fig:SingleArcME}, the procedure for computing approximation error (model cost) for data point $\mathbf{P}_i$ is explained. Since three points define a unique circle, the center of the arc can be simply computed using two arc nodes and one middle node geometrically. One can then calculate the radius of the arc by measuring the distance between any of the three points and the arc center. Next, using the estimated arc radius $r_{est}$, we define a virtual point $\mathbf{P}_i^{v}$ on the arc, as shown in figure \ref{fig:SingleArcME}. Among the points on the arc, $\mathbf{P}_i^{v}$ is chosen so that it lies also on the line that connects the data point $\mathbf{P}_i$ and the arc center $\mathbf{X}_{C}$. Then the arc measurement residual $\mathbf{r}^{i}_{\mathrm{ME}}$ for the $i$th data point is set as the difference between $\mathbf{P}_i^{v}$ and $\mathbf{P}_i$. The algebraic derivation is shown below.

\begin{equation}
\begin{aligned}
    \mathbf{r}^{i}_{\mathrm{ME}}= & \mathbf{P}_i^{v}-\mathbf{P}_i \\
                                = & \mathbf{X}_{c} + r_{est} \frac{\mathbf{P}_{i} - \mathbf{X}_{c}}{\lVert \mathbf{P}_{i} - \mathbf{X}_{c} \rVert} - \mathbf{P}_i \\
                                = & \left( \frac{r_{est}}{\lVert \mathbf{P}_{i} - \mathbf{X}_{c} \rVert} - 1 \right)   \left( \mathbf{P}_{i} - \mathbf{X}_{c} \right)
\end{aligned}
\label{eq:single_ME1}
\end{equation}

\noindent 
In equation \ref{eq:single_ME1}, $r_{est}$ (estimated arc radius), $\mathbf{X}_{c}$ are computed using two arc nodes and the middle node. Finally, with the derived residual $\mathbf{r}^{i}_{\mathrm{ME}}$ for each data point $\mathbf{P}_i$, we can compute the cost function for the arc measurement model as follows.

\begin{equation}
\begin{aligned}
    \mathcal{L}_{\mathrm{ME}} = & \sum_{i=1}^{n}{\lVert \mathbf{r}^{i}_{\mathrm{ME}} \rVert}^{2}_{\Sigma^{i}_{\mathrm{ME}}} \\
    & \sum_{i=1}^{n}{\lVert  \mathbf{P}_i^{v}-\mathbf{P}_i \rVert}^{2}_{\Sigma^{i}_{\mathrm{ME}}} \\
    = & \sum_{i=1}^{n}{\lVert \mathbf{r}^{i}_{\mathrm{ME}} \left( \mathbf{A}_{1}, \mathbf{A}_{2}, \mathbf{N}_{1}, \mathbf{P}_i \right)\rVert}^{2}_{\Sigma^{i}_{\mathrm{ME}}}
\end{aligned}
\label{eq:single_ME2}
\end{equation}

\noindent
Since the virtual point $\mathbf{P}_i^{v}$ is derived from $\mathbf{A}_{1}, \mathbf{A}_{2}, \mathbf{N}_{1}$ and point $\mathbf{P}_i$, we can write the residual $\mathbf{r}^{i}_{\mathrm{ME}}$ as a function of $\mathbf{A}_{1}, \mathbf{A}_{2}, \mathbf{N}_{1}$ and point $\mathbf{P}_i$ in equation \ref{eq:single_ME2}. Here, the squared Mahalanobis Distance is used again for weighting each residual with covariance matrix $\Sigma^{i}_{\mathrm{ME}}$. Also, note that all the data points have different covariance matrices $\Sigma^{i}_{\mathrm{ME}}$, and therefore the arc will be optimized so that approximation error can be reduced further for data points with higher reliability.

\subsubsection{Single Arc Equality Constraint 1: Middle Node}
\label{Chap2EqConst}
The final model included in the cost function is an equality constraint that restricts the relative positions of optimization variables $\mathbf{A}_{1}, \mathbf{A}_{2}, \mathbf{N}_{1}$. Other than the two arc nodes that represent both ends of the arc, we have set the middle point of the arc as one of the arc parameters (optimization variable). This means that the middle node $\mathbf{N}_{1}$ should lie on the perpendicular bisector of line segment $\overline{\mathbf{A}_{1}\mathbf{A}_{2}}$. This can be implemented by taking the inner product of vectors $\overrightarrow{\mathbf{A}_{1}\mathbf{A}_{2}}$ and $\overrightarrow{\mathbf{M}\mathbf{N}_1}$ in figure \ref{fig:SingleArcME}, and equating the result to zero. 

\begin{equation}
    \mathbf{r}_{\mathrm{Eq1}} = {\left( \mathbf{A}_{2} - \mathbf{A}_{1} \right)}^{\top} \left( \mathbf{N}_{1} - \frac{1}{2}\left( \mathbf{A}_{1} + \mathbf{A}_{2}\right) \right)
\label{eq:single_CS1}
\end{equation}

\noindent
In equation \ref{eq:single_CS1}, the dimension of equality constraint is 1(scalar). The equality constraint will be added to the original cost function together with the Lagrange multiplier during optimization.

\subsection{Single Arc Approximation: Augmented Cost Function and Optimization}
\label{Chap2Full}
Wrapping up the proposed cost function models and equality constraint model, we can rewrite the optimization problem as follows.
\begin{equation}
\begin{aligned}
    \min_{\mathbf{A}_{1},\mathbf{A}_{2},\mathbf{N}_{1}} \! \mathcal{L} & = \mathcal{L}_{\mathrm{AC}} + \mathcal{L}_{\mathrm{ME}} \\
    & = {\lVert \mathbf{P}_{1} - \mathbf{A}_{1} \rVert}^{2}_{\Sigma_{\mathrm{AC}}} + {\lVert \mathbf{P}_{n} - \mathbf{A}_{2} \rVert}^{2}_{\Sigma_{\mathrm{AC}}} \\ 
    & +  \sum_{i=1}^{n}{\lVert \mathbf{r}_{\mathrm{ME}} \left( \mathbf{A}_{1}, \mathbf{A}_{2}, \mathbf{N}_{1}, \mathbf{P}_i \right)\rVert}^{2}_{\Sigma^{i}_{\mathrm{ME}}}\\
    \textrm{s.t.} \quad & \mathbf{r}_{\mathrm{Eq1}} = 0
\end{aligned}
\label{eq:single_full}
\end{equation}

\noindent
Equation \ref{eq:single_full} is the final cost function with an equality constraint, and the optimization variables are the two arc nodes $\mathbf{A}_{1}, \mathbf{A}_{2}$ and the middle node $\mathbf{N}_{1}$. This type of optimization problem can be classified as a typical constrained nonlinear least squares(CNLS) optimization problem. 

\hfill \break
\noindent \textbf{Remarks on Model Cost Balancing}

In equation \ref{eq:single_full}, we can observe that augmenting the anchor model and the arc measurement model forms the complete cost function for single arc optimization. Since the anchor model cost and the arc measurement model cost are designed differently, it is important to balance the costs of the two models. Without adding more weights in equation \ref{eq:single_full}, cost balancing can be controlled by tuning the covariance used in each model. However, covariance data used in the arc measurement model are pre-computed(fixed) values, which means that only the anchoring model covariance is available for tuning.

As mentioned in the anchor model (figure \ref{fig:SingleArcAC1}), increasing the diagonal elements of the covariance matrix $\Sigma_{\mathrm{AC}}$ lowers the "weight" of the anchor model. By reducing the weight of the anchor model, it can be interpreted that the weight of the arc measurement model has naturally increased, even if the cost of the arc measurement model remains unchanged when calculating the augmented cost function. Qualitatively, this also means that we are mainly focusing on optimizing the arc shape, rather than putting more emphasis on fixing the arc's end points to the first and last data points tightly. The same way of "weighting" will be performed for chapter \ref{Chap3}.

\subsubsection{Typical Method of Solving Nonlinear Least Squares (NLS) Problem}
Before solving the constrained version of nonlinear least squares, we first introduce how to obtain the optimal solution of unconstrained nonlinear least squares. Among the algorithms introduced in \cite{Madsen2004}, the simplest Gauss-Newton will be explained.

\hfill \break
\noindent \textbf{Unconstrained Nonlinear Least Squares}

Let $\mathbf{f}: \mathbb{R}^n \mapsto \mathbb{R}^m$ be a vector function with $m \geq n$. The main objective is to minimize $\lVert \mathbf{f} \rVert$, or equivalently to find 

\begin{equation}
    \mathbf{x}^{*} = \mathrm{argmin}_{\mathrm{x}} ~ F\left(\mathbf{x}\right)
\label{eq:NLS1}
\end{equation}
\noindent
where

\begin{equation}
F(\mathbf{x})=\frac{1}{2} \sum_{i=1}^m\left(f_i(\mathbf{x})\right)^2=\frac{1}{2}\|\mathbf{f}(\mathbf{x})\|^2=\frac{1}{2} \mathbf{f}(\mathbf{x})^{\top} \mathbf{f}(\mathbf{x})
\label{eq:NLS2}
\end{equation}

For detailed explanation of the Gauss-Newton method, formulas for derivatives of $F$ are derived. Assuming that $\mathbf{f}$ has continuous second partial derivatives, its Taylor expansion can be written as

\begin{equation}
\mathbf{f(x+h) = f(x) + J(x)h} + O \left( {\lVert h \rVert}^2 \right)
\end{equation}

\noindent
where $\mathbf{J} \in \mathbb{R}^{m\times n}$ is referred to as the Jacobian of $\mathbf{f}$. This is a matrix that has the first partial derivatives of the function elements,

\begin{equation}
    {\left( \mathbf{J(x)} \right)}_{ij} = \frac{\partial f_i}{\partial x_j}(\mathbf{x})
\end{equation}

\noindent
And for $F : \mathbb{R}^n \mapsto \mathbb{R}$, the gradient is

\begin{equation}
    \frac{\partial F}{\partial x_j}(\mathbf{x})=\sum_{i=1}^m f_i(\mathbf{x}) \frac{\partial f_i}{\partial x_j}(\mathbf{x}) 
\end{equation}

\noindent
which can be written as

\begin{equation}
    \mathbf{F}^{'}(\mathbf{x}) = \mathbf{J(x)}^{\top} \mathbf{f(x)}
\end{equation}

\hfill \break
\noindent \textbf{Gauss-Newton Method}

For small $\lVert \mathbf{h} \rVert$, we can linearize the vector function $\mathbf{f}$ as follows.

\begin{equation}
    \mathbf{f(x+h)} \simeq l(h) \equiv \mathbf{f(x) + J(x)h}
\end{equation}

\noindent
Then inserting this to equation \ref{eq:NLS2},

\begin{equation}
\begin{aligned}
    F\mathbf{(x+h)} \simeq L\mathbf{(h)} & \equiv \frac{1}{2} l(h)^{\top}l(h) \\
    & = \frac{1}{2} \mathbf{f(x)^{\top}f(x)} + \mathbf{h^{\top}{J(x)}^{\top}f(x)} + \frac{1}{2}\mathbf{h^{\top}{J(x)}^{\top}J(x)h} \\
    & = F\mathbf{(x)} + \mathbf{h^{\top}{J(x)}^{\top}f(x)} + \frac{1}{2}\mathbf{h^{\top}{J(x)}^{\top}J(x)h}
\end{aligned}
\label{eq:NLS3}
\end{equation}

\noindent
From equation \ref{eq:NLS3}, the \textit{Gauss-Newton step} $\mathbf{h}_{\mathrm{gn}}$ minimizes $L(\mathbf{h})$,

\begin{equation}
    \mathbf{h}_{\mathrm{gn}} = \mathrm{argmin}_{\mathbf{h}} ~ L(\mathbf{h})
\end{equation}

\noindent
Also from equation \ref{eq:NLS3}, the gradient and the Hessian of $L$ can be derived.

\begin{equation}
    \mathbf{L^{'}(h) = {J(x)}^{\top}f(x) + {J(x)}^{\top}J(x)h, \quad\quad L^{''}(h) = {J(x)}^{\top}J(x)}
\end{equation}

\noindent
$\mathbf{L}^{''}(\mathbf{h})$ is symmetric and it is also positive definite if $\mathbf{J(x)}$ has full rank. This implies that $L(\mathbf{h})$ has a unique minimizer, which can be found by solving $\mathbf{L^{'}(h)} = \mathbf{0}$.

\begin{equation}
\left(\mathbf{{J(x)}^{\top}J(x)}\right) \mathbf{h}_{\mathrm{gn}} = -\mathbf{{J(x)}^{\top} f(x)} 
\label{eq:NLS4}
\end{equation}

\noindent
The above equation is equivalent to solving a linear equation $\mathbf{Ax = b}$. Large-scale, sparse linear equations can be solved efficiently by using the Column Approximate Minimum Degree Ordering(COLAMD) algorithm \cite{Davis2004}.  The obtained step $\mathbf{h}_{\mathrm{gn}}$ is a descent direction for $F$ since

\begin{equation}
{\mathbf{h}_{\mathrm{gn}}}^{\top} \mathbf{F^{'}(x)} = {\mathbf{h}_{\mathrm{gn}}}^{\top} \mathbf{{J(x)}^{\top}f(x)} = - {\mathbf{h}_{\mathrm{gn}}}^{\top} \left(\mathbf{{J(x)}^{\top}J(x)}\right) \mathbf{h}_{\mathrm{gn}} < 0
\end{equation}

\noindent
Therefore, at each iteration of optimization, the optimization variable (vector) $\mathbf{x}$ is updated as

\begin{equation}
    \mathbf{x}^{k+1} = \mathbf{x}^{k} + \alpha \mathbf{h}^{k}_{\mathrm{gn}}
\end{equation}

\noindent
The equation above describes the variable update at $k$th iteration. Here, $\alpha$ is the step size, which is set as $1$ in the classical Gauss-Newton method. For other advanced methods, various line search methods are used for finding the value of $\alpha$. Other than the Gauss-Newton method, Levenberg-Marquardt method, and Powell's Dog-Leg method are most widely used when solving general unconstrained NLS problems. The core difference between these three algorithms lies in how the update vector $\mathbf{h}$ is calculated.

\subsubsection{Solving Constrained Nonlinear Least Squares (CNLS) Problem}
Equality or inequality constraints are added to the previously introduced NLS to form the CNLS problem. For solving CNLS, advanced techniques such as Barrier / Penalty method, Broyden - Fletcher - Goldfarb - Shanno(BFGS) Hessian Approximation and Lagrange Multiplier are needed. 

Returning to our original problem introduced in equation \ref{eq:single_full}, optimization variables for single arc approximation are the two arc nodes $\mathbf{A}_{1}, \mathbf{A}_{2}$ and the middle node $\mathbf{N}_{1}$. These arc parameters are augmented as a column vector $\mathbf{x}$, and will be iteratively updated in the CNLS solver. The optimal solution of the proposed CNLS problem in equation \ref{eq:single_full} is obtained using 'lsqnonlin.m' of MATLAB optimization toolbox \cite{OptToolDoc}. 

\subsection{Single Arc Approximation: Examples}
Before moving on to multiple arc approximation, we test the proposed single arc approximation with generated data points and covariance. For data generation, white Gaussian noise was added to true points on the arc. The covariance matrix for each data point $\Sigma^{i}_{\mathrm{ME}}$ was set to have random diagonal elements from $1^2$ to $30^2$.

\begin{figure}[!th]
    \centering
    \includegraphics[width=0.65\columnwidth]{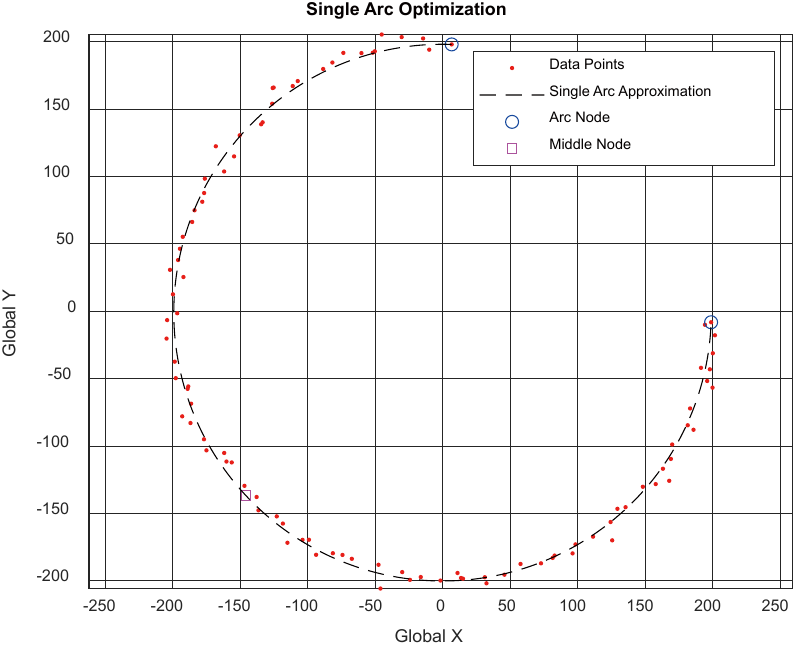}
    \caption{Single Arc Approximation Example 1 (Generated Data Points)}
    \label{fig:SingleArcEx1}
\end{figure}

As we can observe from figure \ref{fig:SingleArcEx1}, noisy generated data points with varying covariance are well-fitted into a single arc. Other than generated data, the single arc approximation is also tested with real-world collected data points. 

\begin{figure}[!th]
    \centering
    \includegraphics[width=\columnwidth]{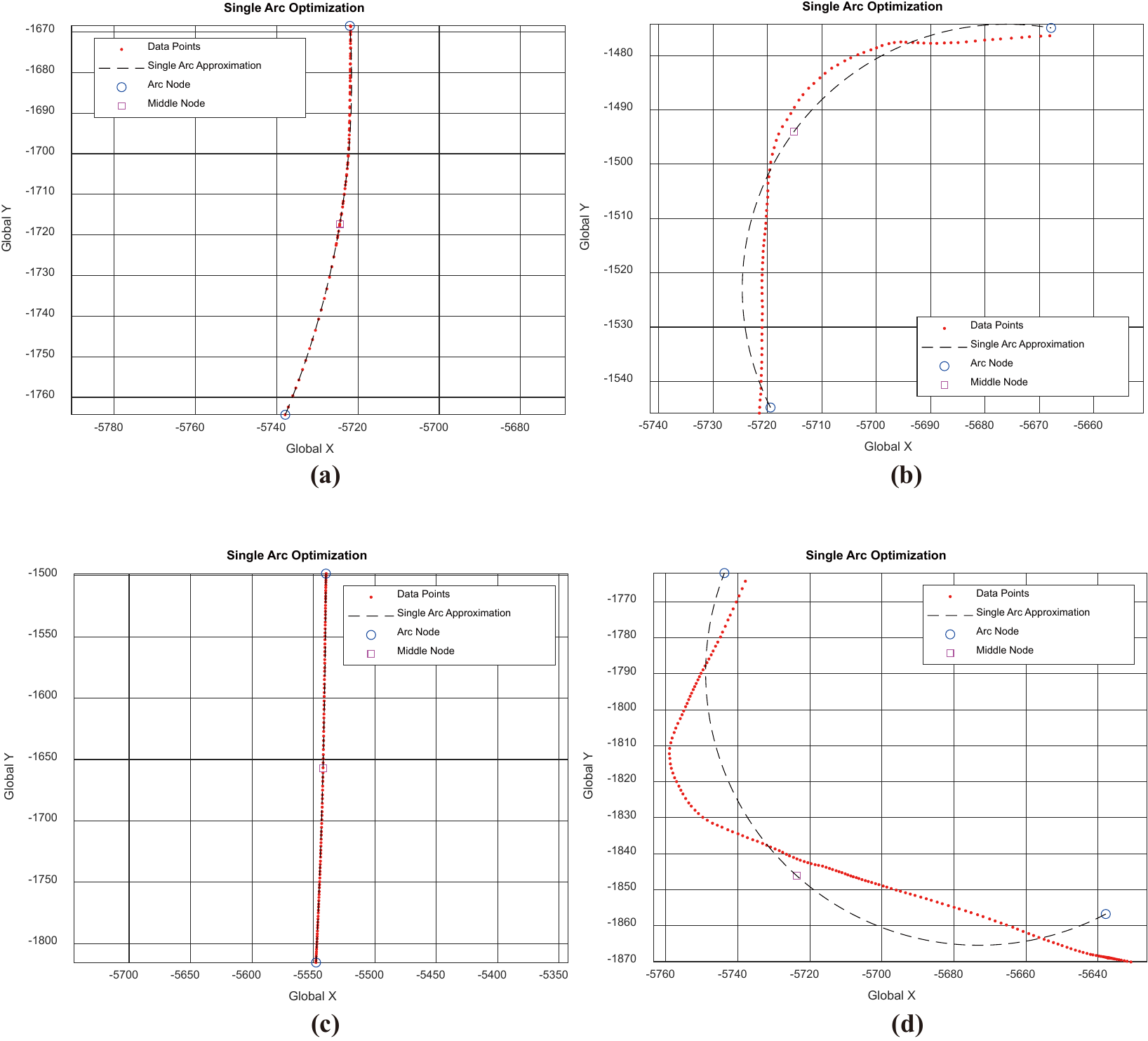}
    \caption{Single Arc Approximation Example 2 (Real-World collected data points from vehicle experiment in Sejong city, South Korea)}
    \label{fig:SingleArcEx2}
\end{figure}

Data points introduced in figure \ref{fig:SingleArcEx2} are computed by fusing vehicle trajectory and ane detection results, in Sejong city, South Korea. The detailed process of obtaining data point covariance will be introduced in chapter \ref{Chap3} and \ref{Appx1}. While single arc approximation seems to be acceptable for cases (a) and (c), it is quite obvious that data approximation with only one arc is not enough for cases (b) and (d). In order to tackle the limits of single-arc approximation, reliability-based multiple-arc approximation will be covered in chapter \ref{Chap3}. 

%
%
%
% Chapter 3
%
%
\newpage
\section{Multiple Arc Approximation}
\label{Chap3}
In this chapter, we extend the concept of the single-arc approximation to multiple-arc approximation. Although the idea seems straightforward, there are several more key factors to consider, as shown below.
\begin{itemize}
    \item Parameters of arc segments should be initialized for stable convergence.
    \item Arc nodes overlap for adjacent arc segments.
    \item (Data point - arc segment) matching is needed for optimization.
    \item All arc segments should satisfy $G^{1}$ continuity.
    \item A validating process of arc parameters is needed.
    \item A determination process of when to end the approximation is needed.
\end{itemize}

The multiple-arc approximation framework will be designed in a way that reflects all the arguments mentioned above. Modified cost functions / constraints will be explained, and the proposed framework will be tested on real-world collected data points and covariance.

\subsection{Multiple Arc Approximation Framework}

\begin{figure}[t!]
    \centering
    \includegraphics[width=0.75\columnwidth]{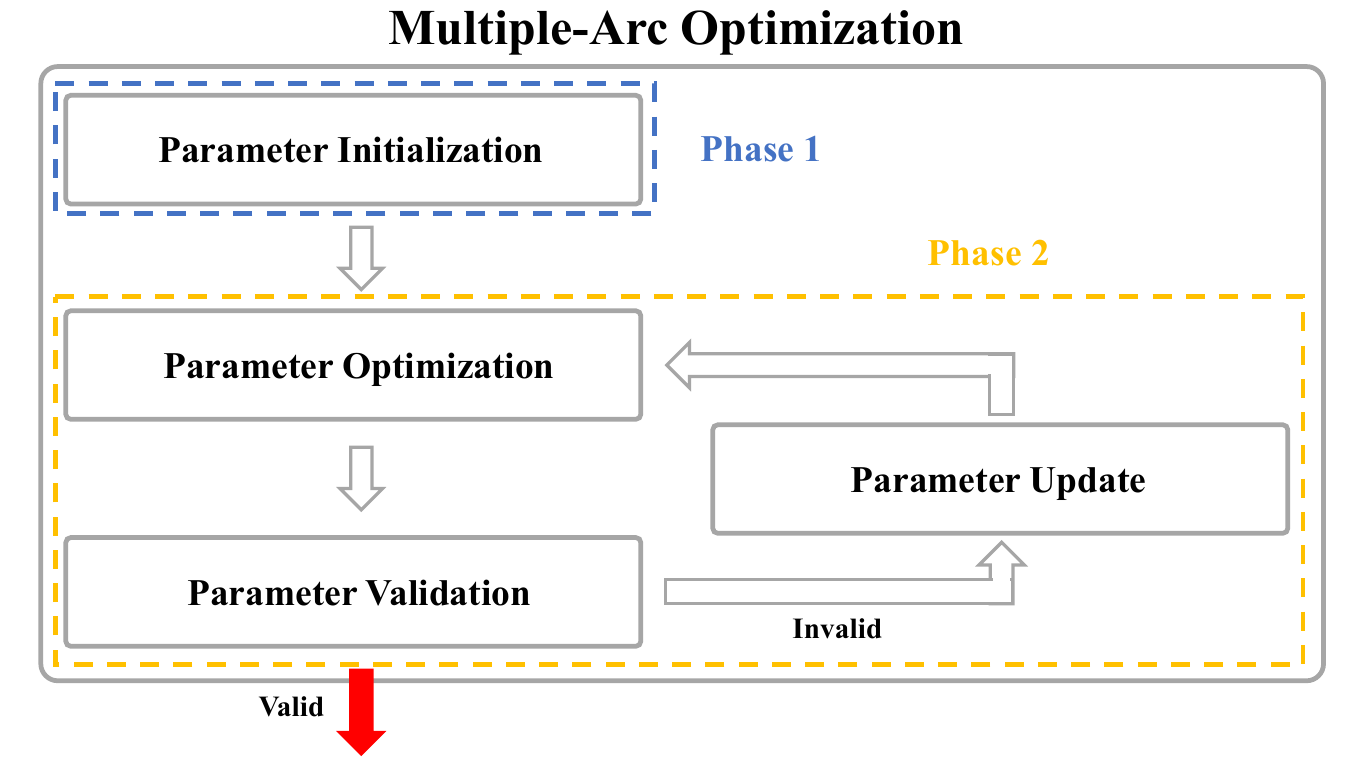}
    \caption{Multiple-Arc Approximation framework can be divided into two phases: Phase 1 is for initialization of various arc nodes and middle nodes, and Phase 2 is for finding a set of arc parameters that satisfies the approximation error condition.}
    \label{fig:MultipleArcFrm}
\end{figure}

The overall framework for multiple-arc approximation is presented in figure \ref{fig:MultipleArcFrm}. The data approximation process can be divided into 2 phases. 

In \textbf{phase 1}, the initial number of arc segments is determined, and corresponding arc parameters (arc nodes and middle nodes) are initialized by utilizing single-arc approximation discussed in chapter \ref{Chap2}. The purpose of initialization is to obtain initial parameter values of adequate quality to avoid divergence during the optimization step in phase 2.

Then in \textbf{phase 2}, CNLS optimization is performed based on several cost function and constraint models. The main difference between single-arc and multiple-arc approximation occurs directly after the arc parameter optimization. While the single-arc approximation ends right away, the multiple-arc approximation framework performs additional arc parameter validation using arc approximation errors and covariance of each data point. If the current arc parameter (arc nodes and middle nodes) set is acceptable after the validation process, optimization ends. If not, the number of segments is increased by one and the parameter optimization step is repeated until all the arc segments are valid. 

\subsection{Multiple Arc Approximation Phase 1: Parameter Initialization}
\label{Chap3Init}
In the phase 1: parameter initialization step, the initial number of arc segments needed is computed. Using this information, the arc parameters are initialized separately using single-arc approximation proposed in chapter \ref{Chap2}. 

\subsubsection{Recursive Linear Approximation of Data Points }
\label{Chap3.2.1}

\begin{algorithm}[t]
\caption{Divide and Conquer based Recursive Linear Approximation}\label{alg:cap}
\begin{algorithmic}
    \State lb $\gets 1$, \quad ub $\gets n $ (Data Length), \quad $\mathbf{P}$: Data Points
    \State Intvs = $\mathbf{LinearApproximation} \left( \mathbf{P}, \: \mathrm{lb}, \: \mathrm{ub} \right)$
    \State \Return Intvs
    \\\hrulefill
    \Function{\textbf{LinearApproximation}}{$\mathbf{P}, \: \mathrm{lb}, \: \mathrm{ub}$}
    \State $(P_x, P_y)$: Full Data Points (X, Y coordinates)
    \State lb: Data Lower Bound, \quad ub: Data Upper Bound
    \State D = zeros(1,ub-lb+1) : Zero-Initialized Row Vector
    \\
    \State m = $\frac{P_y (\mathrm{ub}) - P_y (\mathrm{lb})}{P_x (\mathrm{ub}) - P_x (\mathrm{lb})} $
    \State n = $P_y (\mathrm{lb}) - m \: P_x (\mathrm{lb})$
    
    \For{$i=\mathrm{lb}:\mathrm{ub}$}
        \State $\mathrm{D}(i-\mathrm{lb}+1) = \frac{\left| m \: {P_x} (i) - {P_y} (i) \: + \:n \right|}{\sqrt{m^2\: + \:1}}$
    \EndFor
    \State [maxD,Idx] = min(D)
    \\
    \If{$\mathrm{maxD} > \epsilon$}
        \State $\mathrm{I_1}$ = \textbf{LinearApproximation} $\left( \mathbf{P},\: \mathrm{lb},\:  \mathrm{Idx} \right)$ 
        \State $\mathrm{I_2}$ = \textbf{LinearApproximation} $\left( \mathbf{P},\: \mathrm{Idx},\:  \mathrm{ub} \right)$ 
        \State \Return $[\mathrm{I_1},\: \mathrm{I_2}]$
    \Else
        \State \Return $[\mathrm{lb}, \mathrm{ub}]$
    \EndIf
\EndFunction
\end{algorithmic}
\label{algo:multi}
\end{algorithm}

To determine the initial number of arc segments for approximating data points, the rough shape of the given points should be known. Assuming that the points are well ordered, we perform recursive linearization to approximate data points into polylines(i.e. multiple connected lines). Here, the \textbf{divide-and-conquer} algorithm is implemented for the recursive data point linearization.

A brief explanation of algorithm \ref{algo:multi} is as follows. We assume that there are a total of $n$ data points. 
\begin{enumerate}[Step 1.]
    \item Set initial interval of interest as [1, \textrm{n}]
    \item Connect the first index and the last index data point with a single line.
    \item Find the index of the point(\textrm{Idx}) with the largest linear fit error.
    \item Check if the largest error is above the threshold (Yes/No).
    \item 
    \begin{enumerate}
        \item (Yes) Divide intervals into [1, \textrm{Idx}], [\textrm{Idx}, \textrm{n}] and repeat from Step 2.
        \item (No) Return current intervals (Will be propagated back)
    \end{enumerate}
\end{enumerate}

\begin{figure}[!ht]
    \centering
    \includegraphics[width=\columnwidth]{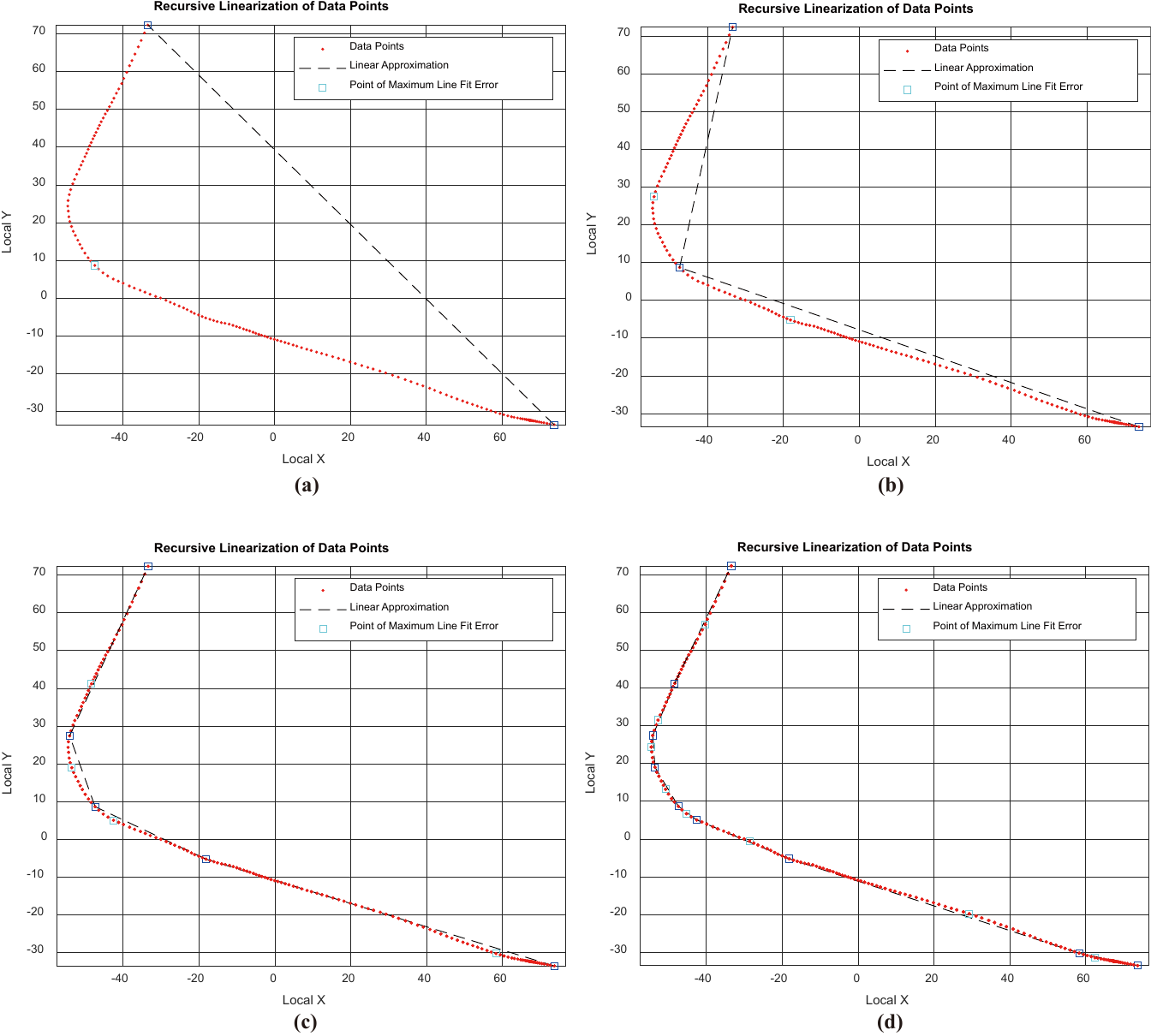}
    \caption{Full Process of Recursive Linearization: Tested on sample data points (from (a) to (d))}
    \label{fig:MultipleArcReLin}
\end{figure}
A sample result of recursive linear approximation is shown in figure \ref{fig:MultipleArcReLin}. As a result of recursive linear approximation, data point intervals for piecewise linear approximation can be obtained. Since these linear approximation intervals contain the rough shape of given data points, we can now determine the initial number of arc segments(i.e. the number of arc nodes and middle nodes needed for creating the initial optimization variable).

\subsubsection{Determining Initial Number of Arc Segments}
\label{Chap3.2.2}
\begin{figure}[!th]
    \centering
    \includegraphics[width=0.75\columnwidth]{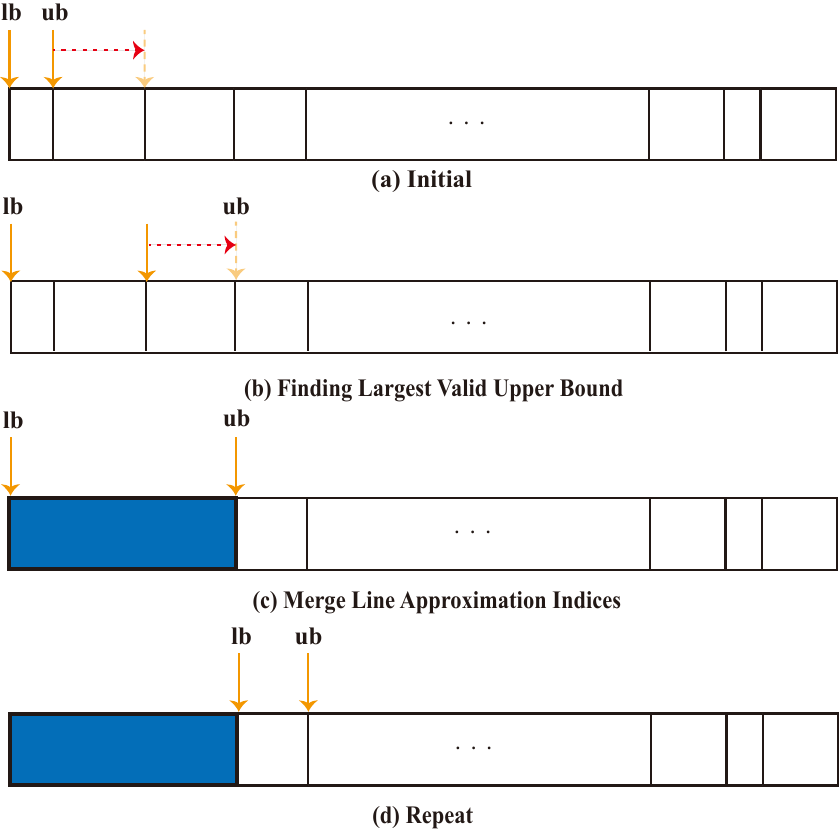}
    \caption{Merging Linear Approximation Intervals for Single-Arc Approximation: \textbf{lb} and \textbf{ub} stands for data lower and upper bound respectively. At step (b), the data upper bound index is increased along the linear approximation interval boundaries. This is repeated until single-arc approximation is invalid for data points within the index range of \textbf{lb} and \textbf{ub}. If the single-arc approximation is invalid for current \textbf{ub}, \textbf{ub} is reverted to the most recent interval boundary. Then at step (c), the data indices from \textbf{lb} and \textbf{ub} are merged, meaning that these data points will be approximated as a single arc at the initialization phase. Finally, for step (d), steps (b) to (c) are repeated until \textbf{ub} reaches the end.}
    \label{fig:MultipleArcMerge}
\end{figure}
% figure
The basic idea of determining the initial number of arc segments is to incrementally merge linear approximation intervals obtained previously in \ref{Chap3.2.1} and then test single-arc approximation validity. The process of merging the linear approximation interval is introduced graphically in figure \ref{fig:MultipleArcMerge}. The vertical separation lines represent linear approximation index interval boundaries. At step (b), as the upper bound data index \textbf{ub} is updated as the next boundary index, the single-arc approximation is performed for data points between the index of \textbf{lb} to \textbf{ub}. This process is repeated while single-arc approximation results are acceptable. On the other hand, when the data approximation is invalid, the upper bound index is reverted to the most recently valid boundary index. At step (c) of figure \ref{fig:MultipleArcMerge}, we can conclude that data points between the index of \textbf{lb} to \textbf{ub} can be approximated by a single arc. The validation method of single-arc approximation will be discussed in the future section(Phase 2: Ending Criterion). After (c), the lower bound index is moved to the latest upper bound index to re-initialize \textbf{lb} and \textbf{ub}. Steps (b) to (c) are repeated until the upper bound index reaches the final data point. In this way, the initial number of arc segments and the initial arc approximation intervals needed to represent given data points can be computed.

\subsubsection{Multiple Arc Parameter Initialization}
\begin{figure}[!th]
    \centering
    \includegraphics[width=0.5\columnwidth]{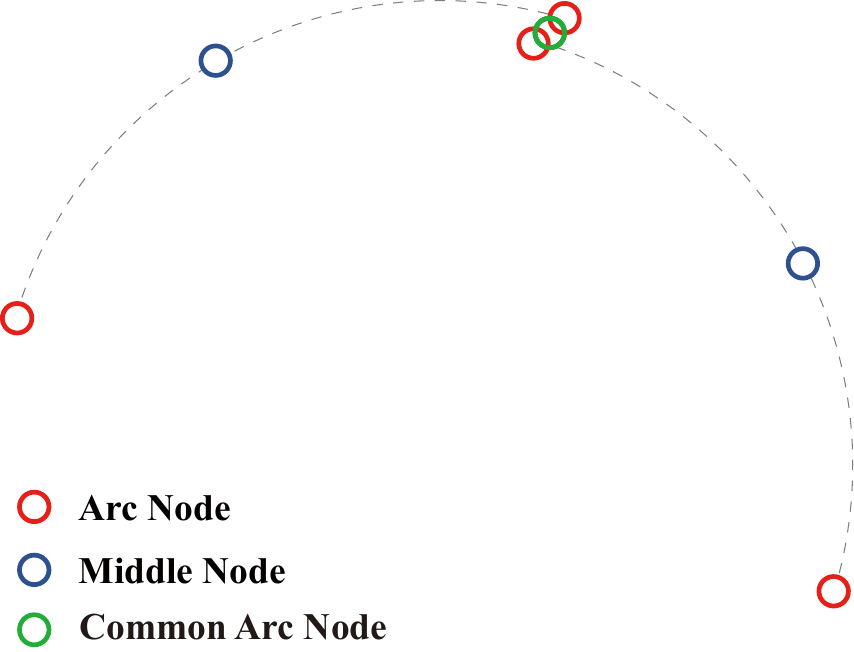}
    \caption{Parameter Initialization for Multiple Arcs: In an example with two adjacent arc segments, the optimization variable is initialized so that the two arcs can be optimized simultaneously. To consider two arcs as a single set of arc parameters, the common arc node (green point) is defined to be shared between two arcs.}
    \label{fig:MultipleArcParams}
\end{figure}
Since the initial intervals for multiple arc approximation are obtained in chapter \ref{Chap3.2.2}, the arc parameters (arc nodes and middle nodes) for each separate arc segment can be initialized via single-arc approximation. We perform single-arc approximation for each initial multiple-arc approximation interval and save (2 arc nodes, 1 middle node) for each arc segment. However, as mentioned at the very start of this chapter, we have to consider that more than one arc is to be parameterized simultaneously. For instance, if two arc segments are to be connected as shown in figure \ref{fig:MultipleArcParams}, they should share one arc node at the connection region, rather than having two separate arc nodes. Therefore, to handle overlapping arc nodes for adjacent arc segments, we allocated one common arc node for two overlapping arc nodes. 

The position of the common arc node (colored green) is initialized by taking the mean of two overlapping arc nodes' positions. In figure \ref{fig:MultipleArcParams} for example, a total of 3 arc nodes(2 original arc nodes + 1 common arc node) and 2 middle nodes will be augmented as the initial optimization variable. Even for general cases where there are more than two arc segments, common arc nodes can be initialized similarly. After parameter initialization in phase 1, the initialized arc nodes(including common arc nodes) and the middle nodes will be optimized based on the cost/constraint models for multiple-arc approximation.

\subsection{Multiple Arc Approximation Phase 2: Parameter Optimization}
\label{Chap3Opt}
Moving on to multiple-arc approximation framework phase 2, as shown in figure \ref{fig:MultipleArcFrm}, the initialized arc parameters from phase 1 will be optimized with modified cost function models and constraints, and will also be validated using arc approximation error and each data point's covariance matrix. If all the arc segments are acceptable after evaluation, the optimization loop ends. On the other hand, if there are some invalid arc segments, the number of arc segments is increased by one, and the optimization loop is repeated. Here, note that the number of arc segments is fixed within the arc parameter optimization process (left top block of phase 2 in figure \ref{fig:MultipleArcFrm}). 

Focusing on cost function models, slight modifications were made to the original cost function models and equality constraint introduced in chapter \ref{Chap2}. Moreover, 2 more constraint models were added due to the properties of the multiple-arc approximation framework. 

\subsubsection{Data Association}
\label{Chap3DA}
\begin{figure}[t]
    \centering
    \includegraphics[width=0.9\columnwidth]{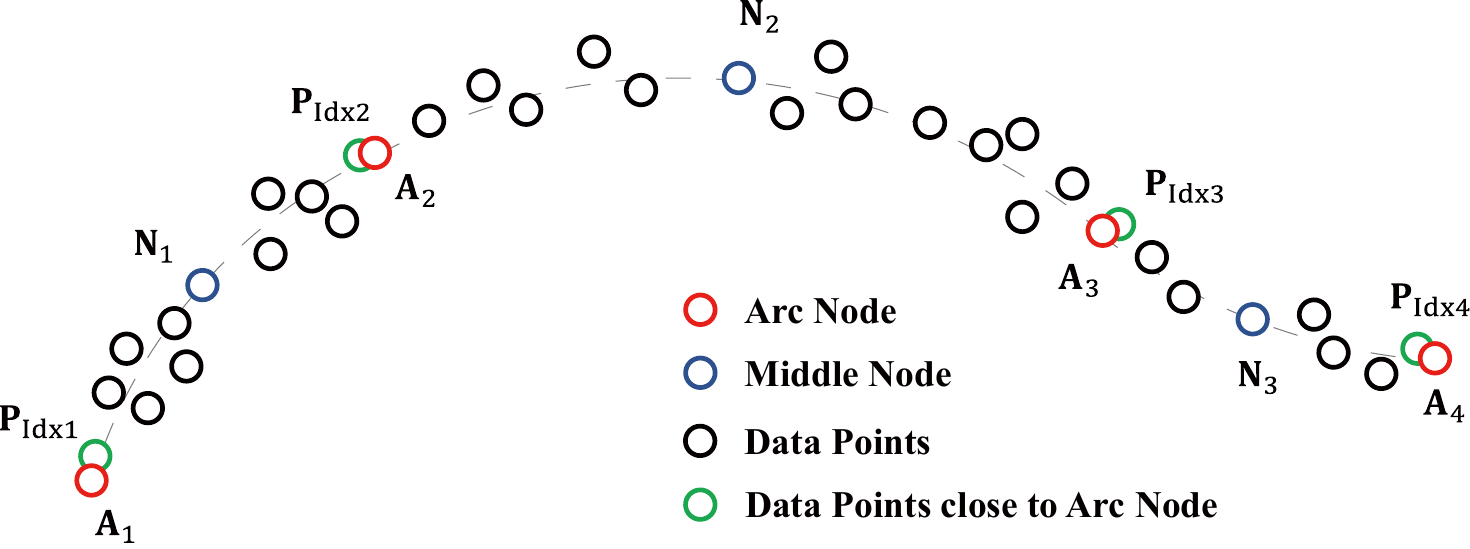}
    \caption{Data Association for 3 Arcs: (1) Find the closest data point to each arc node (marked green) (2) Data points between index \textrm{Idx} $i$ to \textrm{Idx} $i+1$ are matched to arc segment number $i$. For example, points between $\mathbf{P}_{\mathrm{Idx1}}$ and $\mathbf{P}_{\mathrm{Idx2}}$ are matched to the first arc segment. (3) After each optimization iteration, the optimization variables $\mathbf{A}_{1}$, $\mathbf{A}_2$, and so forth, are updated. Subsequently, the association process from step (1) to step (2) is reiterated.}
    \label{fig:MultipleArcDA}
\end{figure}

Before moving on to the cost function and constraint model explanation, we first handle data association, which is the process of matching data points and arc segments. Unlike single-arc approximation, where all the data points are matched to one arc, the matching relationship between data points and multiple arc segments may become ambiguous during the multi-arc approximation. Therefore, for a particular arc segment, we need to decide which data points are going to be matched to the arc segment in the data association step. For example in figure \ref{fig:MultipleArcDA}, since there are 3 arc segments, we need to divide data points into 3 groups during data association. Assuming that the data points are well-sorted, fast data association can be performed by using the index of data points (will be written as \textbf{Idx}) that are closest to arc nodes. In the case of starting ($\mathbf{A}_{1}$) and ending ($\mathbf{A}_{4}$) arc nodes, the first and the last data points will be used (Index number 1 and $n$) respectively. As a result, as shown in figure \ref{fig:MultipleArcDA}, data points that have indices between \textbf{Idx(1)} (= 1) and \textbf{Idx(2)} are linked to the first arc segment. This is the same for the remaining two arc segments. Other than the example addressed previously, the same logic can be applied to various cases with different numbers of arc segments.

The result of data association will be used in the extended cost function and constraint models. Note that data association matching results may critically affect the approximation performance, especially in the arc measurement model. Approximation using wrong data points will lead to convergence failure, instability, and large approximation errors. 

\hfill \break
\noindent \textbf{Remarks on Data Association}

In phase 2 parameter optimization step, the positions of multiple arc nodes and middle nodes are changed for each optimization iteration. Since data association is done based on the positions of arc nodes, it is necessary to perform data association after every optimization iteration.    

\subsubsection{Anchor Model}
In comparison to the single-arc approximation, there are several arc nodes to anchor in the multiple-arc approximation framework. Using the results of data association, the arc nodes and data points can be matched for anchoring. For example, in figure \ref{fig:MultipleArcDA}, arc node $\mathbf{A}_1$ will be matched to data point $\mathbf{P}_{\mathrm{Idx}1}$, and in general, arc node $\mathbf{A}_{i}$ will be matched to data point $\mathbf{P}_{\mathrm{Idx}(i)}$. 

Considering that the initial and final arc nodes are consistently associated with the first and last data points in the data association process, it is rational to assign small values to the diagonal elements of the anchor model covariance. This effectively constrains the first and the last arc nodes, ensuring they remain close to their corresponding matched data points.

Conversely, the data association results for the remaining arc nodes may exhibit inconsistency. This is due to the iterative modification of arc node positions during the optimization step, potentially causing variations in the index of the closest data point. Therefore large values were assigned to the diagonal elements of the anchor model covariance for the remaining arc nodes. This will allow these arc nodes to freely "explore" the solution space for stable convergence. 

Assuming there are $m$ arc segments, then there will be $m+1$ arc nodes ($\mathbf{A}_1, \cdots, \mathbf{A}_{m+1}$). Then the anchor model cost function can be written as follows. 

\begin{equation}
    \mathcal{L}_{\mathrm{AC}} = {\lVert \mathbf{P}_{\mathrm{Idx}(1)} - \mathbf{A}_{1} \rVert}^{2}_{\Sigma_{\mathrm{AC_{1}}}} + {\lVert \mathbf{P}_{\mathrm{Idx}(m+1)} - \mathbf{A}_{m+1} \rVert}^{2}_{\Sigma_{\mathrm{AC_{1}}}} + \sum_{i=2}^{m} {\lVert \mathbf{P}_{\mathrm{Idx}(i)} - \mathbf{A}_{i} \rVert}^{2}_{\Sigma_{\mathrm{AC_{2}}}}
\label{eq:multiple_AC}
\end{equation}

\noindent
In the above equation, $\mathbf{P}$ indicates the data point vector and $\mathbf{Idx}(i)$ is the data point index obtained from data association (\ref{Chap3DA}). In particular, $\mathbf{Idx}(1) = 1$ and $\mathbf{Idx}(m+1) = n$, where $n$ is the total number of data points. The structure of the anchor model for multiple-arc approximation is identical to the single-arc approximation anchor model: the positional difference between an arc node and the matched data point is weighted by a covariance matrix. The first two terms in equation \ref{eq:multiple_AC} are designed to fix the first and last arc nodes to the first and last data points respectively. The remaining terms are for anchoring the remaining arc nodes to their corresponding data points. As explained previously, small values are assigned to the diagonal terms of $\Sigma_{\mathrm{AC_{1}}}$ and large values are assigned to the diagonal elements of $\Sigma_{\mathrm{AC_{2}}}$. 

\subsubsection{Arc Measurement Model}
For the arc measurement model in multiple-arc approximation, equation \ref{eq:single_ME2} derived in the single-arc approximation is repeatedly computed for multiple arcs. Referring to figure \ref{fig:MultipleArcDA}, when computing arc measurement cost for the second arc segment, arc nodes $\mathbf{A}_2, \mathbf{A}_3$, middle node $\mathbf{N}_2$, and data points from index $\mathbf{Idx}(2)$ to $\mathbf{Idx}(3)$ are used. In general, for arc segment $i$, the arc measurement model cost is computed for data points of index $\mathbf{Idx}(i)$ to $\mathbf{Idx}(i+1)$, using arc nodes $\mathbf{A}_{i}, \mathbf{A}_{i+1}$, and middle node $\mathbf{N}_{i}$. The arc measurement model cost for multiple arcs is written below. 

\begin{equation}
    \mathcal{L}_{\mathrm{ME}} = \sum_{i=1}^{m} {\sum_{j=\mathrm{Idx(i)}}^{\mathrm{Idx(i+1)}}{\lVert \mathbf{r}_{\mathrm{ME}} \left( \mathbf{A}_{i}, \mathbf{A}_{i+1}, \mathbf{N}_{i}, \mathbf{P}_j \right)\rVert}^{2}_{\Sigma^{j}_{\mathrm{ME}}}}
\label{eq:multiple_ME}
\end{equation}

In equation \ref{eq:multiple_ME}, note that the covariance matrix $\Sigma^{j}_{\mathrm{ME}}$ is uniquely defined for each data point (pre-computed).

\subsubsection{Equality Constraint 1: Middle Node}
The equality constraint introduced in the single-arc approximation (\ref{Chap2EqConst}) is expanded to accommodate multiple arc segments. Since the underlying principle for the equality constraint is identical to single-arc approximation, equation \ref{eq:single_CS1} is computed for each arc segment in the multiple-arc approximation framework. Assuming that there are $m$ arc segments, then there will be $m+1$ arc nodes and $m$ middle nodes. Then the residual vector for the middle node equality constraint would have the size of $\mathbb{R}^{m}$. Specifically for arc segment $i$, the related arc parameters are arc nodes $\mathbf{A}_{i}, \mathbf{A}_{i+1}$ and the middle node $\mathbf{N}_{i}$. Ultimately, employing these values allows the formulation of the middle node equality constraint for arc segment $i$ as follows:

\begin{equation}
    \mathbf{r}_{\mathrm{Eq1}}(i) = {\left( \mathbf{A}_{i+1} - \mathbf{A}_{i} \right)}^{\top} \left( \mathbf{N}_{i} - \frac{1}{2}\left( \mathbf{A}_{i} + \mathbf{A}_{i+1}\right) \right), \quad \mathrm{for} \; i=1:m
\label{eq:multiple_CS1}
\end{equation}

\noindent
Note that the column vector $\mathbf{r}_{\mathrm{Eq1}}$ is constrained as a zero vector during the parameter optimization.

\subsubsection{Equality Constraint 2: $G^{1}$ Continuity}
\begin{figure}[!th]
    \centering
    \includegraphics[width=0.7\columnwidth]{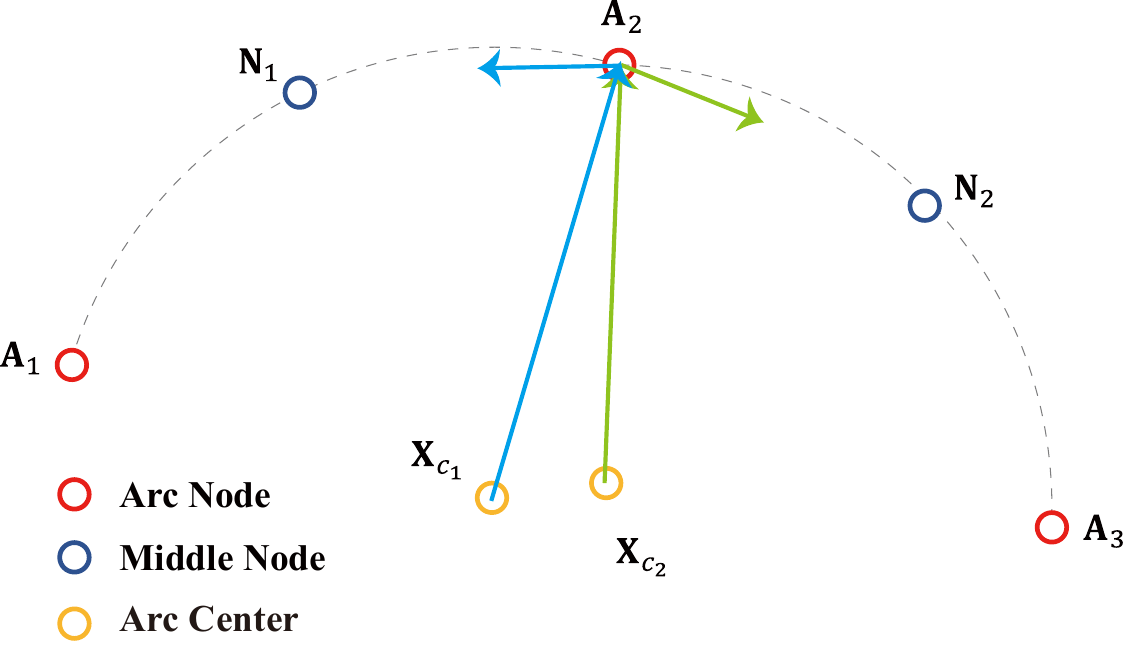}
    \caption{Equality Constraint for $G^{1}$ Continuity}
    \label{fig:MultipleArcCS2}
\end{figure}
If there is more than 1 arc segment, $G^{1}$ continuity between each adjacent arc segment must be satisfied. Figure \ref{fig:MultipleArcCS2} shows that the two adjacent arc segments are not $G^{1}$ continuous. In order for these two segments to be $G^{1}$ continuous, the following conditions should be satisfied. 
\begin{itemize}
    \item (Orthogonality between Blue Vectors) \\ Tangential vector of arc segment 2 at $\mathbf{A}_2$ is orthogonal to vector $\overrightarrow{\mathbf{X}_{c_{1}}\mathbf{A}_2}$ 
    \item (Orthogonality between Green Vectors) \\ Tangential vector of arc segment 1 at $\mathbf{A}_2$ is orthogonal to vector $\overrightarrow{\mathbf{X}_{c_{2}}\mathbf{A}_2}$ 
\end{itemize}

\noindent

Extending the case introduced in figure \ref{fig:MultipleArcCS2}, if we consider the $G^{1}$ continuity constraint between $i$th and $(i+1)$th arc segment, 
arc nodes $\mathbf{A}_{i}, \mathbf{A}_{i+1}$, and $\mathbf{A}_{i+2}$ correspond to $\mathbf{A}_1, \mathbf{A}_2$, and $\mathbf{A}_3$ in figure \ref{fig:MultipleArcCS2} respectively. Then the conditions above can be written into equations as follows. 

\begin{equation}
\begin{aligned}
    &\mathbf{v}_{b_1} = {\left[{\left(\mathbf{A}_{i+1}\right)}_y - {\left(\mathbf{X}_{c_{i}}\right)}_y  ; \; {\left(\mathbf{X}_{c_{i}}\right)}_x - {\left(\mathbf{A}_{i+1}\right)}_x\right]} \\
    &\mathbf{v}_{b_2} = {\left[{\left(\mathbf{A}_{i+1}\right)}_x - {\left(\mathbf{X}_{c_{i+1}}\right)}_x ; \; {\left(\mathbf{A}_{i+1}\right)}_y - {\left(\mathbf{X}_{c_{i+1}}\right)}_y\right]} \\
    &\mathbf{v}_{g_1} = {\left[{\left(\mathbf{A}_{i+1}\right)}_y - {\left(\mathbf{X}_{c_{i+1}}\right)}_y  ; \; {\left(\mathbf{X}_{c_{i+1}}\right)}_x - {\left(\mathbf{A}_{i+1}\right)}_x\right]} \\
    &\mathbf{v}_{g_2} = {\left[{\left(\mathbf{A}_{i+1}\right)}_x - {\left(\mathbf{X}_{c_{i}}\right)}_x ; \; {\left(\mathbf{A}_{i+1}\right)}_y - {\left(\mathbf{X}_{c_{i}}\right)}_y\right]} \\
    &\mathbf{r}_{\mathrm{Eq2}}(2i-1:2i) = \left[{\mathbf{v}_{b_1}}^{\top}  \mathbf{v}_{b_2}; \quad {\mathbf{v}_{g_1}}^{\top}  \mathbf{v}_{g_2}\right] \; , \mathrm{for} \; i=1:m-1
\end{aligned}
\end{equation} 

\noindent
In the equation above, subscript $\left( \mathbf{v} \right)_{x}$ and $\left( \mathbf{v} \right)_{y}$ refer to the $x$ and $y$ component of vector $\mathbf{v}$ respectively. The inner product ${\mathbf{v}_{b_1}}^{\top}  \mathbf{v}_{b_2}$ represents the orthogonality condition between blue vectors and the inner product ${\mathbf{v}_{g_1}}^{\top}  \mathbf{v}_{g_2}$ represents the orthogonality condition between green vectors. For $m$ arc segments, the equality constraint residual vector $\mathbf{r}_{\mathrm{Eq2}}$ has the size of $\mathbb{R}^{2m-2}$ and will be kept as zero vector during optimization. 

\subsubsection{Inequality Constraint 1: Minimum Arc Length}
\begin{figure}[b]
    \centering
    \includegraphics[width=0.7\columnwidth]{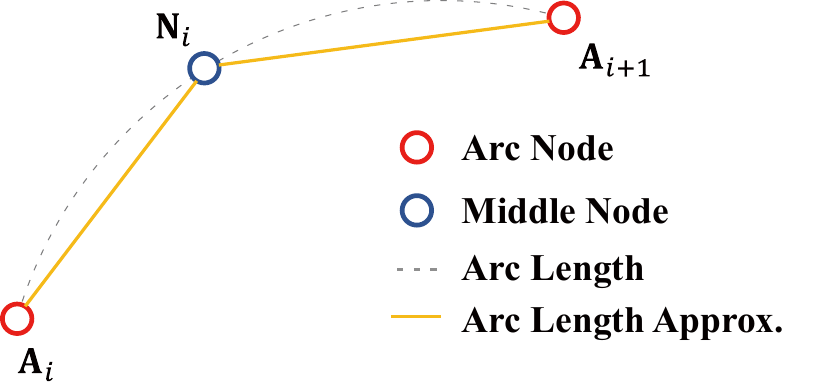}
    \caption{Inequality Constraint for Minimum Arc Length: The true arc length is approximated with the length of orange lines. $\overline{\mathbf{A}_{i} \mathbf{N}_{i}} + \overline{\mathbf{N}_{i} \mathbf{A}_{i+1}} \geq \mathrm{L_{min}}$ is set as the inequality constraint.}
    \label{fig:MultipleArcCS3}
\end{figure}

The final constraint model enforces the arc segments to have a minimum length of $\mathrm{L_{min}}$. The model aims to prevent arc segments from collapsing to a single point(i.e. 2 arc nodes and the middle node converging to the same point), which causes singularity problems during the optimization process. 

For example in figure \ref{fig:MultipleArcCS3}, if we compute the true arc length of segment $i$ using $\mathbf{A}_{i}, \mathbf{A}_{i+1}$, and $\mathbf{N}_{i}$, the value would be severely nonlinear. Setting the true arc length to be larger than $\mathrm{L_{min}}$ would cause the CNLS solver to slow down, or even fail in extreme cases. A way around this problem is to find some simple approximation of the arc length. One method is to approximate the arc length by summing the lengths of $\overline{\mathbf{A}_{i} \mathbf{N}_{i}}$ and $\overline{\mathbf{N}_{i} \mathbf{A}_{i+1}}$, as shown in figure \ref{fig:MultipleArcCS3}. Note that the true arc length is always greater than the sum of the length of two line segments $\overline{\mathbf{A}_{i} \mathbf{N}_{i}}$ and $\overline{\mathbf{N}_{i} \mathbf{A}_{i+1}}$ geometrically. Therefore if we set $\lVert \mathbf{A}_{i} - \mathbf{N}_{i} \rVert + \lVert \mathbf{N}_{i} - \mathbf{A}_{i+1} \rVert$ to be larger than $\mathrm{L_{min}}$, the true arc length will be constrained to have larger value than $\mathrm{L_{min}}$. The inequality constraint residual vector of size $\mathbb{R}^{m}$ is written as follows.

\begin{equation}
    \mathbf{r}_{\mathrm{InEq1}}(i) = 1 - \frac{\lVert \mathbf{A}_{i} - \mathbf{N}_{i} \rVert + \lVert \mathbf{N}_{i} - \mathbf{A}_{i+1} \rVert}{\mathrm{L_{min}}} \; , \; \mathrm{for} \; i=1:m
\end{equation}

\noindent
The inequality above is computed for all the arc segments (from $i=1$ to $m$). 
When performing optimization, the inequality constraint residual vector $\mathbf{r}_{\mathrm{InEq1}}$ is constrained to be less than or equal to a 0 vector.

\subsubsection{Multiple Arc Approximation: Augmented Cost Function and Constraints}
Merging the cost function models and equality/inequality constraint models, we obtain the full CNLS problem structure. Assuming we have $m$ arc segments, the augmented cost function can be written as follows.

\begin{equation}
\begin{aligned}
    \min_{\mathbf{A}_{1}, \cdots \mathbf{A}_{m+1},\mathbf{N}_{1},\cdots \mathbf{N}_{m}} \! \mathcal{L} & = \mathcal{L}_{\mathrm{AC}} + \mathcal{L}_{\mathrm{ME}} \\
    & = {\lVert \mathbf{P}_{1} - \mathbf{A}_{1} \rVert}^{2}_{\Sigma_{\mathrm{AC_{1}}}} + {\lVert \mathbf{P}_{n} - \mathbf{A}_{m+1} \rVert}^{2}_{\Sigma_{\mathrm{AC_{1}}}} \\
    & + \sum_{i=2}^{m} {\lVert \mathbf{P}_{\mathrm{Idx}(i)} - \mathbf{A}_{i} \rVert}^{2}_{\Sigma_{\mathrm{AC_{2}}}}\\ 
    & +  \sum_{i=1}^{m} {\sum_{j=\mathrm{Idx(i)}}^{\mathrm{Idx(i+1)}}{\lVert \mathbf{r}_{\mathrm{ME}} \left( \mathbf{A}_{i}, \mathbf{A}_{i+1}, \mathbf{N}_{i}, \mathbf{P}_j \right)\rVert}^{2}_{\Sigma^{j}_{\mathrm{ME}}}}\\
    \textrm{s.t.} \quad & \mathbf{r}_{\mathrm{Eq1}} = \mathbf{0}, \; \mathbf{r}_{\mathrm{Eq2}} = \mathbf{0}, \; \mathbf{r}_{\mathrm{InEq1}} \leq \mathbf{0}
\label{eq:multiple_Full}
\end{aligned}
\end{equation}

Similar to single-arc optimization, the cost function balancing can be done by tuning covariance matrix $\Sigma_{\mathrm{AC_{1}}}, \Sigma_{\mathrm{AC_{2}}}$. Having initialized arc nodes and middle nodes (computed in \ref{Chap3Init}) as the input to the CNLS solver, the output will be the optimized positions of arc nodes and middle nodes. The CNLS problem given as equation \ref{eq:multiple_Full} is solved by using the interior point method \cite{Coleman1994} implemented in 'lsqnonlin.m' of MATLAB optimization toolbox. The mathematical procedure for attaining the optimal solution mirrors the explanation provided in section \ref{Chap2Full}.

\subsection{Multiple Arc Approximation Phase 2: Parameter Validation}
\label{Chap3VA}
Upon revisiting the overall framework (refer to figure \ref{fig:MultipleArcFrm}), the optimization of arc nodes and middle nodes for multiple arcs is conducted using the CNLS solver within the parameter optimization block. However, even after the solver completes the optimization process, the accuracy of the multiple-arc approximation may be poor due to insufficient arc segments, as depicted in (b) and (d) of Figure \ref{fig:SingleArcEx2}. Consequently, each arc segment undergoes analysis to verify the validity of the arc approximation using matched data points in the parameter validation block. If all the arc segments are acceptable, the multiple-arc approximation is finished promptly. Otherwise, the number of arc segments is increased in the parameter update block, and the parameter optimization is reiterated with an additional arc segment, as illustrated in figure \ref{fig:MultipleArcFrm}.

While many curve-fitting/approximation algorithms use simple RMSE for evaluating approximations, naively using the RMSE value is an inappropriate approach if covariance matrices of data points are given. In our research, instead of RMSE, \textbf{Chi-squared}$\left({\chi}^2 \right)$ \textbf{test}\cite{Pearson1900} is conducted for all data points to determine whether the arc approximation of each arc segment is acceptable or not. The validation steps for an arc segment are:

\begin{enumerate}[Step 1.]
    \item For an arc segment, obtain the matched data point indices from the data association process.
    \item Compute the arc measurement residual \ref{Chap2ME} for each data point. For arc segment $i$ and point index $j$, $\mathbf{r}_{\mathrm{ME}} \left( \mathbf{A}_{i}, \mathbf{A}_{i+1}, \mathbf{N}_{i}, \mathbf{P}_j \right)$ is computed.
    \item Normalize the residual computed in Step 2 using the covariance matrix(squared Mahalanobis distance) and test whether this value is larger than the Chi-squared test threshold (Larger/Smaller).
    \item 
    \begin{enumerate}
        \item (Larger) Arc approximation is invalid for the current data point.
        \item (Smaller) Arc approximation is valid for the current data point.
    \end{enumerate}
    \item If the total number of invalid arc approximations exceeds the threshold $N$, the corresponding arc segment is considered invalid.
\end{enumerate}

In the explanation above, if the squared Mahalanobis distance for arc measurement residual is larger than the Chi-squared test threshold, this indicates that approximation using the optimized arc parameters is not valid for the specific data point. Typically, 99\% confidence level is chosen for the Chi-squared test threshold value. 

\begin{figure}[t]
    \centering
    \includegraphics[width=0.9\columnwidth]{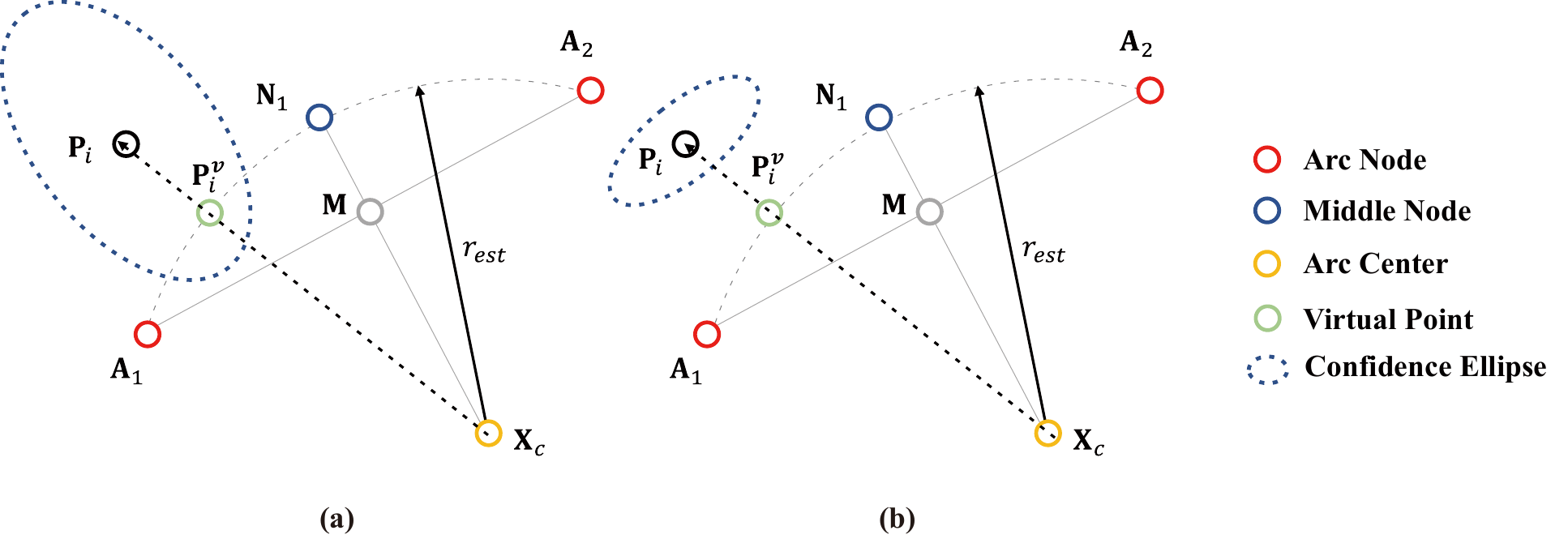}
    \caption{Determining the validity of arc approximation: Assuming that we are performing analysis on arc segment 1 and data point $\mathbf{P}_{i}$, we check if the virtual point $\mathbf{P}^{v}_{i}$(marked green) is inside the confidence ellipse, which is centered at the data point. This validation is done for all the data points and their matched arc segments. Figures (a) Inside: Valid Approximation (b) Outside: Invalid Approximation }
    \label{fig:MultipleArcVA}
\end{figure}

This means that the arc approximation is considered invalid, only if it is placed outside of the 99\% confidence ellipse(drawn using covariance matrix), as shown in figure \ref{fig:MultipleArcVA}.

Ultimately, in order to assess the validity of the approximated arc segment, we count the number of invalid data point approximations for each arc segment. If the total number of invalid data point approximations surpasses the threshold $N$, a tuning variable, the present arc segment is deemed invalid. It is crucial to note that the value of $N$ requires careful tuning. If $N$ is set too high, arc segments with substantial approximation errors might be accepted. Conversely, if $N$ is set too low, a greater number of arc segments may be necessary to accurately approximate data points.

\subsection{Multiple Arc Approximation Phase 2: Parameter Update}
\label{Chap3UP}
The parameter update step is performed if there exists any invalid arc segment after the \textbf{Parameter Validation} (\ref{Chap3VA}) step. In order to prevent the number of arc segments from increasing without bounds, we add one arc segment for every parameter update step. For an effective update, the arc segment with the most number of invalid data point approximations is halved. Original arc nodes and the middle node are used to initialize the arc nodes and the middle node of the newly generated arc segment. 

\begin{figure}[!t]
    \centering
    \includegraphics[width=0.7\columnwidth]{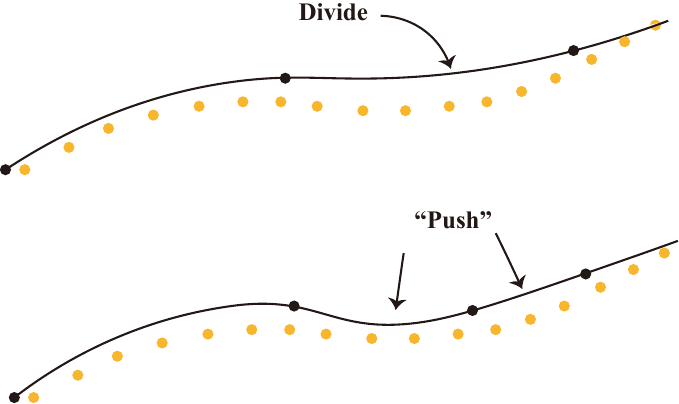}
    \caption{Parameter Update: Divide the most invalid arc segment into half (Top figure). Using the updated parameters as initial optimization variables, the parameter optimization step is performed again to reduce arc approximation error. We can interpret this as "pushing" the approximating curve toward the data points (Bottom figure).}
    \label{fig:MultipleArcUp}
\end{figure}

As shown in figure \ref{fig:MultipleArcUp}, when optimization is performed again after the parameter update, the new arc node will gradually converge to the near local minimum point. Excluding the first and the last arc nodes, since the anchors of the remaining arc nodes are set with anchor model covariance of large diagonal terms, the remaining arc nodes can move rather freely around the anchor point. Moreover, due to the data association step in every optimization iteration, the anchor point itself is shifted towards the data point near the possible local minima.  

\subsection{Multiple Arc Approximation: Examples}
% Things to add 
% 1. Covariance extraction process
% 2. Comparison test with other algorithm
% 3. Limitations of current work? / possible applications

Integrating four different blocks: parameter initialization (\ref{Chap3Init}), parameter optimization (\ref{Chap3Opt}), parameter validation (\ref{Chap3VA}), and parameter update (\ref{Chap3UP}) in the multiple-arc approximation framework, as shown in figure \ref{fig:MultipleArcFrm}, we are ready to evaluate the performance of our proposed algorithm. 

\subsubsection{Covariance Extraction Process for Our Dataset}
\begin{figure}[t]
    \centering
    \includegraphics[width=0.9\columnwidth]{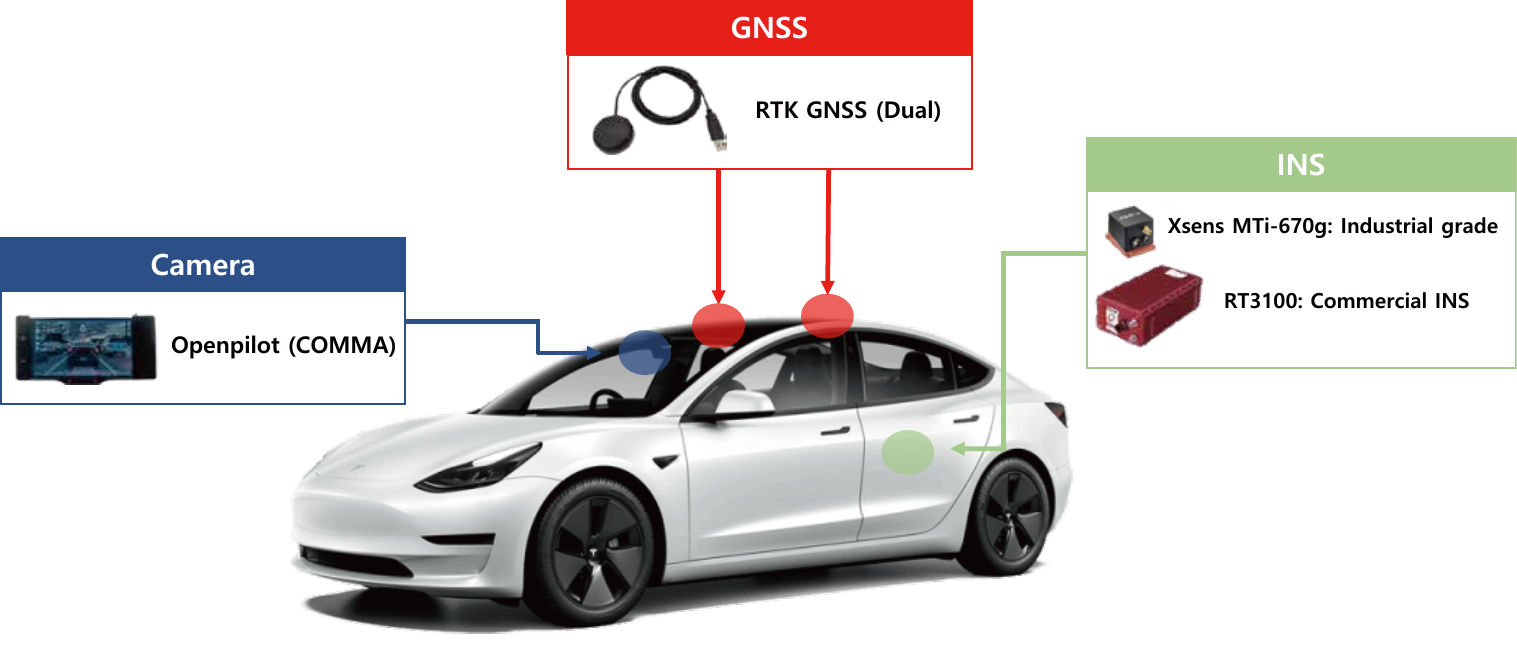}
    \caption{Test Vehicle Sensor Configuration: Lane detection and extraction is performed in OpenPilot \cite{Openpilot}, and vehicle position is estimated by fusing GNSS and INS sensor data.}
    \label{fig:Vehicle}
\end{figure} 

Before proceeding with the evaluation, it's essential to elucidate the process of deriving each data point and its covariance after the vehicle test drive in Sejong City. The test vehicle and sensor configuration are illustrated in figure \ref{fig:Vehicle}. The data points obtained in this study pertain to lane points, which are determined by leveraging information from the vehicle position and lane detection results. The vehicle position is calculated through sensor fusion of the Global Navigation Satellite System (GNSS) and Inertial Navigation System (INS), while the task of lane detection is carried out by OpenPilot \cite{Openpilot}, an Advanced Driver Assistance System (ADAS) software. Additionally, the reliability of lateral lane detection, expressed as standard deviation, is also extracted from OpenPilot. This reliability information is then integrated to derive the covariance for the lane points \cite{Jeon2023}. Further details on this process can be found in \ref{Appx1}.

Prior to the assessment of the proposed multiple-arc approximation algorithm, the real-world lane point data and its associated covariance are obtained through the detailed procedure outlined above.

\subsubsection{Multiple Arc Approximation of Simple Examples}

\begin{figure}[t]
    \centering
    \includegraphics[width=\columnwidth]{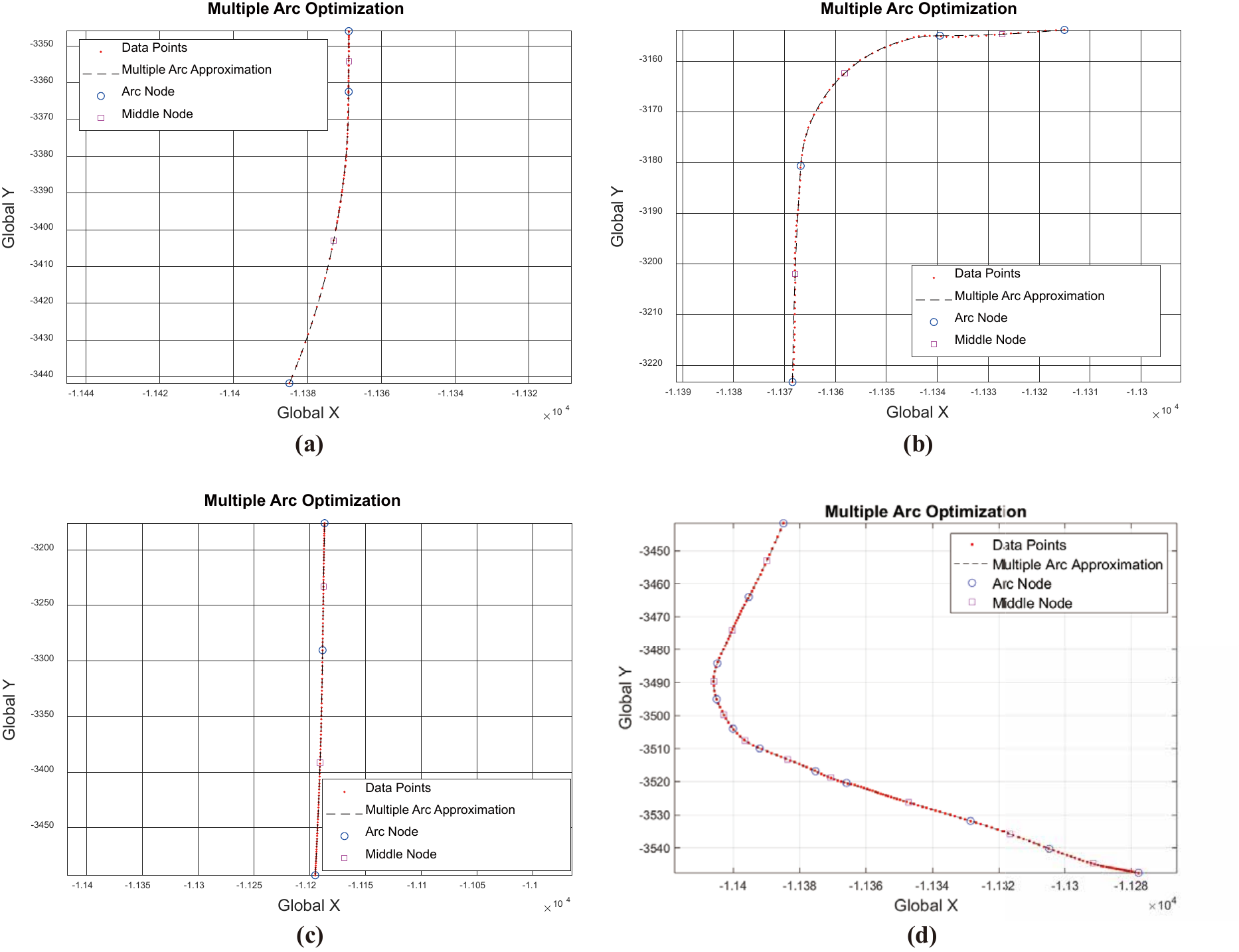}
    \caption{Multiple Arc Optimization Example 1 (Same data points as figure \ref{fig:SingleArcEx2})}
    \label{fig:MultipleArcEx1}
\end{figure}

The multiple-arc optimization framework is employed in the same examples introduced in the single-arc optimization. In figure \ref{fig:MultipleArcEx1}, all the data points undergo optimization with two or more valid arc segments. It's important to note that an arc approximation with a lower Root Mean Square Error (RMSE) doesn't necessarily signify a superior approximation result, as there could be more invalid Chi-squared test samples. Given that covariance matrices for all data points are pre-computed, evaluation based on Chi-squared tests is deemed more reasonable than merely calculating the RMSE for each arc segment.

\subsubsection{Multiple Arc Approximation Application: Lane Map Parameterization}

\begin{figure}[t]
    \centering
    \includegraphics[width=\columnwidth]{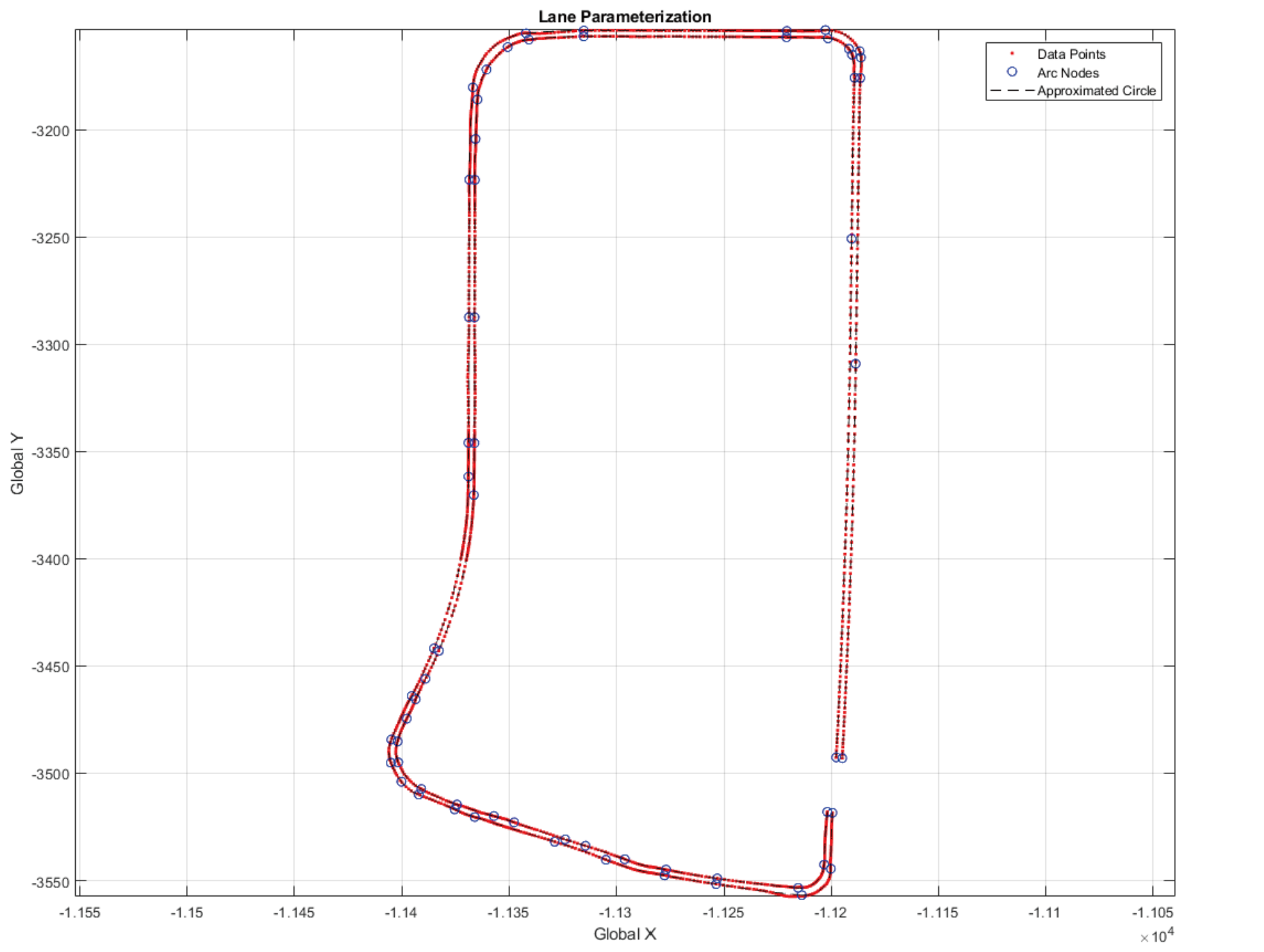}
    \caption{Multiple Arc Optimization Example 2 (Full data points from vehicle trip)}
    \label{fig:MultipleArcEx2}
\end{figure}

Finally, as the last example of multiple-arc optimization, we introduce the multiple-arc approximation results of the left and right ego lanes of a vehicle trip in Sejong City, South Korea. The real-world examples introduced at single-arc optimization and multiple-arc optimization previously in figures \ref{fig:SingleArcEx2} and \ref{fig:MultipleArcEx1} were partially sampled from this whole trip. In figure \ref{fig:MultipleArcEx2}, the left/right lane data points and arc nodes were drawn together, but optimized separately.

\begin{table}[thb]
    \centering
    \begin{tabular}{|c|c|c|}
    \hline
    Direction & Total Arc Segment Length & Total Number of Segments \\
    \hline \hline
    Left & 1101.12 m & 28 \\ 
    \hline
    Right & 1078.37 m & 34 \\
    \hline
    \end{tabular}
    \caption{Summary of Left and Right Ego Lane Arc Parameterization}
    \label{tab:MultipleEx2}
\end{table}

A total of 768 data points were parameterized into several arc segments for each left/right ego lane. A summary of multiple-arc optimization results is listed in table \ref{tab:MultipleEx2}. If we analyze the results, 768 data points from the left ego lane can be simply represented with 57 control points(i.e. 28 middle nodes + 29 arc nodes = 57 control points) and the right ego lane can be represented with 69 control points without disobeying the reliability conditions(Chapter \ref{Chap3VA}).

\subsubsection{Comparison}
\begin{figure}
    \centering
    \includegraphics[width=0.95\columnwidth]{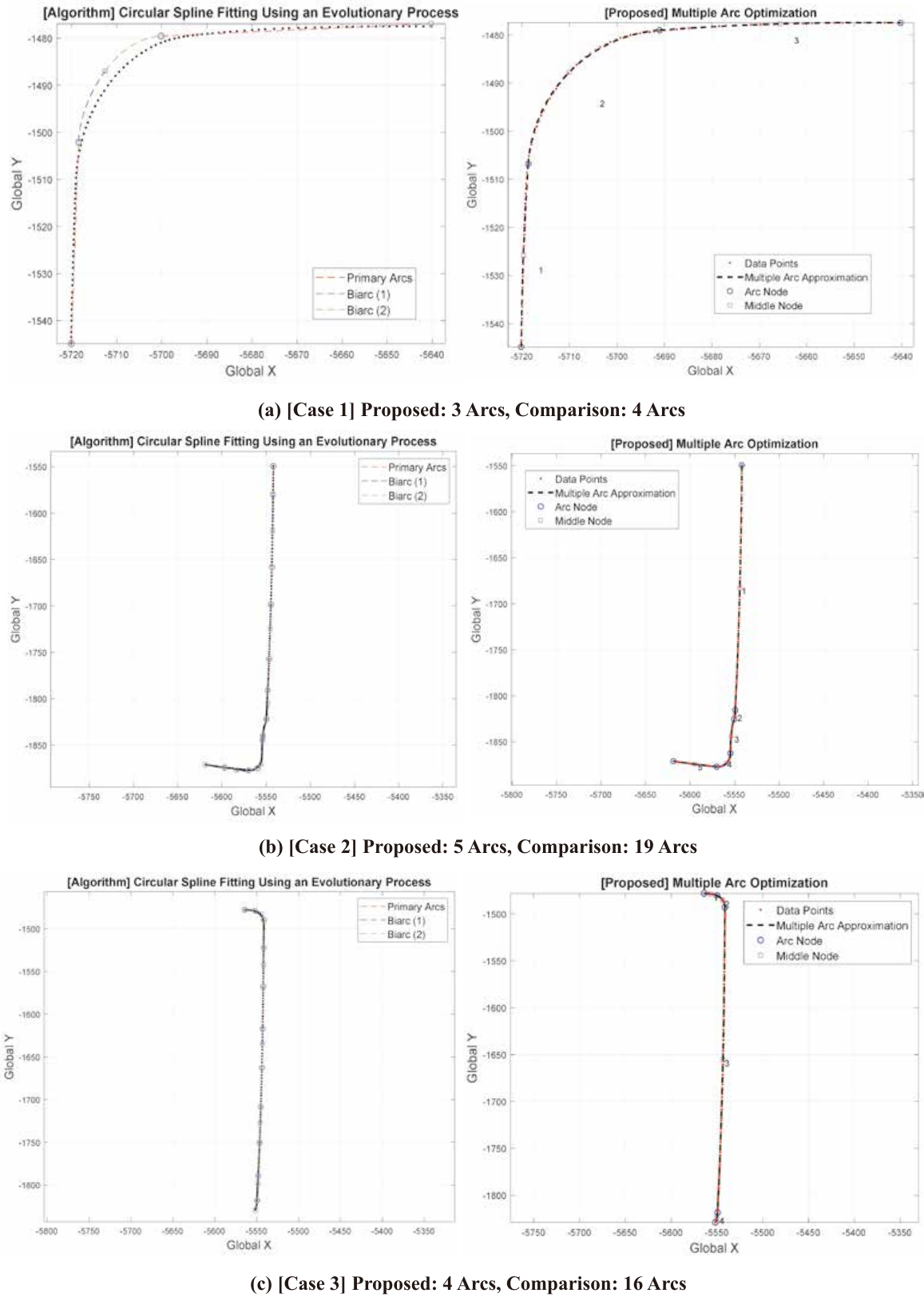}
    \caption{Comparison of \cite{Song2009} and Our Proposed Framework}
    \label{fig:Comparison}
\end{figure}

For a fair evaluation of our proposed framework, we also compare our work with the arc approximation proposed in \cite{Song2009}. The original algorithm in \cite{Song2009} handles 3D problems, but it was re-implemented to suit our 2D data approximation problem. As explained in the introduction chapter, authors of \cite{Song2009} approximated data points by connecting several primary arcs with biarcs. As shown in figure \ref{fig:Comparison}, for a total of 3 example data point sets collected in Sejong City, we tested our proposed multiple-arc approximation framework, together with the reproduced work of \cite{Song2009} for performance comparison.

From figure \ref{fig:Comparison}, we can observe that our algorithm outperforms the work of \cite{Song2009} by reducing the number of arcs needed to approximate the same set of data points. Moreover, the algorithm in \cite{Song2009} shows instability if the density of data points varies throughout the dataset. Since raw data points are sampled for initial arc generation in \cite{Song2009}, it is quite obvious that a dataset with considerable noise will cause serious problems in the comparison algorithm. The concept of employing biarcs for arc spline approximation is interesting, but as explained above, it is unstable for general usage and also not appropriate for compact data approximation.

\subsubsection{Analysis}
% Possible limitations
% Advantages
Before concluding our research, we analyze the advantages and the possible limitations of our reliability-based arc spline approximation framework.

\hfill \break
\noindent \textbf{Advantages}

\begin{itemize}
    \item Robust to noisy data points
    \item Compact data approximation by multiple arcs
\end{itemize}

\noindent \textbf{Limitations}

\begin{itemize}
    \item Data points should be well-ordered (sorted)
    \item Covariance for data points should be accurate
    \item No guarantee of returning the minimum number of arc segments
    \item Optimized arc parameters may be sub-optimal solutions
\end{itemize}

The first and second drawbacks of the proposed framework can be mitigated by other pre-processing algorithms. For example, sorting unorganized points can be done by applying the 'moving least squares method' introduced in \cite{LEE2000}. Moreover, the third and fourth limitations can be resolved by adding the '0-norm' minimization\cite{Lima2015} to our proposed framework, resulting in a Multi-Objective Optimization (MOO) problem. 

\section{Conclusion}
\label{Chap4}
In this research, novel optimization frameworks for both single and multiple arc approximations are proposed. While conventional arc spline approximation algorithms focus on minimizing simple RMSE, the proposed method finds a set of statistically optimal arc parameters, given data points and their corresponding covariance matrices. Evaluation of the proposed optimization framework was performed using various real-world collected data points and showed successful results.  

As shown in real-world examples, a possible extension to this research would be applying the multiple-arc approximation framework to vehicle lane mapping. Since current digital maps store lane data in the format of points and line segments, we believe our proposed method will show great efficiency and effectiveness in data storage and manageability. Moreover, compared to conventional arc spline methods, lane segment information can be updated after data collection from overlapping trips, by using the proposed, reliability-based method. Therefore in future research, multiple-arc approximation will be performed and evaluated for multiple vehicle trips in larger regions of Sejong City.


\appendix

\section{Lane Point Covariance Extraction}
\label{Appx1}
The lane point can be obtained via the vector equation below. 

\begin{equation}
\begin{aligned}
& \boldsymbol{\mathrm{p}}^{L}_{i} = \boldsymbol{\mathrm{p}}_i +  \mathrm{R}_{i} \boldsymbol{L}_{i} \\
& \boldsymbol{\mathrm{p}}^{R}_{i} = \boldsymbol{\mathrm{p}}_i +  \mathrm{R}_{i} \boldsymbol{R}_{i}
\end{aligned}
\label{eq:cov_ext1}
\end{equation}

In equation \ref{eq:cov_ext1}, $\mathrm{p}^{L}_{i}, \mathrm{p}^{R}_{i}$ indicate the left and right lane data point, and $\boldsymbol{L}_{i}, \boldsymbol{R}_{i}$ stands for the left and right lane measurement centered at the $i$th vehicle state respectively. Moreover, $\mathrm{R}_{i}$ refers to the rotational matrix for describing vehicle attitude (roll, pitch, yaw angles). After obtaining the covariance for vehicle states using \cite{Sparseinv}, we can approximate the 2D covariance of the lane points($\mathrm{p}^{L}_{i}, \mathrm{p}^{R}_{i}$) by applying the Multivariate-$\delta$ method\cite{MultiDelta}, as shown below. 

\scriptsize
\begin{equation}
\operatorname{Var}\left(\boldsymbol{\mathrm{p}}^{L}_{i}\right) = \left[\begin{array}{ccc}
\frac{\partial \boldsymbol{\mathrm{p}}^{Lx}_{i}}{\partial \delta \boldsymbol{\mathrm{p}}_i} & \frac{\partial \boldsymbol{\mathrm{p}}^{Lx}_{i}}{\partial \delta {\phi}_i } & \frac{\partial \boldsymbol{\mathrm{p}}^{Lx}_{i}}{\partial \delta \boldsymbol{L}_i}  \\
\frac{\partial \boldsymbol{\mathrm{p}}^{Ly}_{i}}{\partial \delta \boldsymbol{\mathrm{p}}_i} & \frac{\partial \boldsymbol{\mathrm{p}}^{Ly}_{i}}{\partial \delta {\phi}_i } & \frac{\partial \boldsymbol{\mathrm{p}}^{Ly}_{i}}{\partial \delta \boldsymbol{L}_i}  \\
\frac{\partial \boldsymbol{\mathrm{p}}^{Lz}_{i}}{\partial \delta \boldsymbol{\mathrm{p}}_i} & \frac{\partial \boldsymbol{\mathrm{p}}^{Lz}_{i}}{\partial \delta {\phi}_i } & \frac{\partial \boldsymbol{\mathrm{p}}^{Lz}_{i}}{\partial \delta \boldsymbol{L}_i}  
\end{array}\right]\left[\begin{array}{ccc}
\operatorname{Var}\left(\delta \boldsymbol{\mathrm{p}}_i\right) & \cdots & \operatorname{Cov}\left(\delta \boldsymbol{\mathrm{p}}_i, \delta \boldsymbol{L}_i\right) \\
\vdots & \ddots & \vdots \\
\operatorname{Cov}\left(\delta \boldsymbol{\mathrm{p}}_i, \delta \boldsymbol{L}_i\right) & \cdots & \operatorname{Var}\left( \delta \boldsymbol{L}_i\right)
\end{array}\right]\left[\begin{array}{ccc}
\frac{\partial \boldsymbol{\mathrm{p}}^{Lx}_{i}}{\partial \delta \boldsymbol{\mathrm{p}}_i} & \frac{\partial \boldsymbol{\mathrm{p}}^{Lx}_{i}}{\partial \delta {\phi}_i } & \frac{\partial \boldsymbol{\mathrm{p}}^{Lx}_{i}}{\partial \delta \boldsymbol{L}_i}  \\
\frac{\partial \boldsymbol{\mathrm{p}}^{Ly}_{i}}{\partial \delta \boldsymbol{\mathrm{p}}_i} & \frac{\partial \boldsymbol{\mathrm{p}}^{Ly}_{i}}{\partial \delta {\phi}_i } & \frac{\partial \boldsymbol{\mathrm{p}}^{Ly}_{i}}{\partial \delta \boldsymbol{L}_i}  \\
\frac{\partial \boldsymbol{\mathrm{p}}^{Lz}_{i}}{\partial \delta \boldsymbol{\mathrm{p}}_i} & \frac{\partial \boldsymbol{\mathrm{p}}^{Lz}_{i}}{\partial \delta {\phi}_i } & \frac{\partial \boldsymbol{\mathrm{p}}^{Lz}_{i}}{\partial \delta \boldsymbol{L}_i} 
\end{array}\right]^{\top}
\label{eq:MultiDelta}
\end{equation}
\normalsize

\noindent
The second matrix refers to the covariance of vehicle states and lateral lane point measurement. Assuming we have already computed this matrix, we only need to compute the first matrix (Jacobian matrix) to obtain the covariance of lane points. In the equation \ref{eq:MultiDelta}, the $\delta$ term in front of variables means an infinitesimal change in the variable. For the case of $\delta {\phi}_i$, it indicates the infinitesimal angular change of the rotational matrix $\mathrm{R}_{i}$. The partial derivatives in the Jacobian matrix are derived algebraically as follows. 

\noindent
\textbf{(1) Vehicle Position}

\begin{equation}
\begin{aligned}
\boldsymbol{\mathrm{p}}^{L}_{i} \left( \delta \boldsymbol{\mathrm{p}}_i \right) & = \boldsymbol{\mathrm{p}}_i +  \mathrm{R}_{i} \boldsymbol{L}_{i} + \mathrm{R}_{i} \delta \boldsymbol{\mathrm{p}}_i\\
& = \boldsymbol{\mathrm{p}}^{L}_{i} + \mathrm{R}_{i} \delta \boldsymbol{\mathrm{p}}_i \qquad \qquad \therefore \frac{\partial \boldsymbol{\mathrm{p}}^{L}_{i}}{\partial \delta \boldsymbol{\mathrm{p}}_i} = \mathrm{R}_i
\end{aligned}
\label{eq:vehicle_delta}
\end{equation}

\noindent
\textbf{(2) Rotational Matrix}
\begin{equation}
\begin{aligned}
\boldsymbol{\mathrm{p}}^{L}_{i} \left( \delta  {\phi}_i \right) & = \boldsymbol{\mathrm{p}}_i +  \mathrm{R}_{i} \operatorname{Exp} \left( \delta {\phi}_i \right) \boldsymbol{L}_{i} \\
& \cong  \boldsymbol{\mathrm{p}}_i + \mathrm{R}_{i} \left( \mathbf{I} + {\delta \phi}^{\wedge}_{i} \right) \boldsymbol{L}_{i} \\
& = \boldsymbol{\mathrm{p}}_i + \mathrm{R}_{i} \boldsymbol{L}_{i} + \mathrm{R}_{i} {\delta \phi}^{\wedge}_{i} \boldsymbol{L}_{i} \\
& = \boldsymbol{\mathrm{p}}^{L}_{i} + \mathrm{R}_{i} {\delta \phi}^{\wedge}_{i} \boldsymbol{L}_{i} \\ 
& = \boldsymbol{\mathrm{p}}^{L}_{i} - \mathrm{R}_{i} \boldsymbol{L}^{\wedge}_{i} {\delta \phi}_{i} \qquad \qquad  \therefore \frac{\partial \boldsymbol{\mathrm{p}}^{L}_{i}}{\partial \delta {\phi}_i} \cong -\mathrm{R}_i \boldsymbol{L}^{\wedge}_{i}
\end{aligned}
\label{eq:rot_delta}
\end{equation}

\noindent
\textbf{(3) Lane Measurement}

\begin{equation}
\begin{aligned}
\boldsymbol{\mathrm{p}}^{L}_{i} \left( \delta \boldsymbol{L}_{i} \right) & = \boldsymbol{\mathrm{p}}_i +  \mathrm{R}_{i} \boldsymbol{L}_{i} + \mathrm{R}_{i} \delta \boldsymbol{L}_i \\
& = \boldsymbol{\mathrm{p}}^{L}_{i} + \mathrm{R}_{i} \delta \boldsymbol{L}_i \qquad \qquad \therefore \frac{\partial \boldsymbol{\mathrm{p}}^{L}_{i}}{\partial \delta \boldsymbol{L}_i} = \mathrm{R}_i
\end{aligned}
\label{eq:lane_delta}
\end{equation}

\noindent
The $\mathrm{Exp}$ term in the equation above refers to the exponential mapping of $\mathbb{R}^{3}$ vector to $SO(3)$ (lie algebra), and $\wedge$ is the skew-symmetric transform operator for $\mathbb{R}^{3}$ vector. Equation \ref{eq:vehicle_delta}, \ref{eq:rot_delta}, \ref{eq:lane_delta} finds the first, second, and third column of the Jacobian matrix in equation \ref{eq:MultiDelta} respectively. The notation for equations \ref{eq:vehicle_delta} $\sim$ \ref{eq:lane_delta} is borrowed from \cite{Forster2015}.

\section*{Acknowledgement}
This research was supported by the BK21 FOUR Program of the National Research Foundation Korea(NRF) grant funded by the Ministry of Education(MOE); Autonomous Driving Technology Development Innovation Program (20018181, Development of Lv. 4+ autonomous driving vehicle platform based on point-to-point driving to logistic center for heavy trucks) funded by the Ministry of Trade, Industry \& Energy(MOTIE, Korea) and Korea Evaluation Institute of Industrial Technology(KEIT); Technology Development for Future Automotives Tuning Parts (P0021036, Development of a smart damper system for electric vehicles capable of tuning damping force while driving) funded by the Ministry of Trade, Industry \& Energy (MOTIE, Korea) and the Korea Institute for Advancement of Technology (KIAT); the Technology Innovation Program(20014983, Development of autonomous chassis platform for a modular vehicle)funded By the Ministry of Trade, Industry \& Energy(MOTIE, Korea);
\bibliographystyle{elsarticle-num}
\bibliography{reference}

\begin{thebibliography}{10}
\expandafter\ifx\csname url\endcsname\relax
  \def\url#1{\texttt{#1}}\fi
\expandafter\ifx\csname urlprefix\endcsname\relax\def\urlprefix{URL }\fi
\expandafter\ifx\csname href\endcsname\relax
  \def\href#1#2{#2} \def\path#1{#1}\fi

\bibitem{MEEK1995}
D.~Meek, D.~Walton, Approximating smooth planar curves by arc splines, Journal
  of Computational and Applied Mathematics 59~(2) (1995) 221--231.
\newblock \href {https://doi.org/https://doi.org/10.1016/0377-0427(94)00029-Z}
  {\path{doi:https://doi.org/10.1016/0377-0427(94)00029-Z}}.

\bibitem{Jee2003}
S.~Jee, T.~Koo, Tool-path generation for nurbs surface machining, in:
  Proceedings of the 2003 American Control Conference, 2003., Vol.~3, 2003, pp.
  2614--2619 vol.3.
\newblock \href {https://doi.org/10.1109/ACC.2003.1243471}
  {\path{doi:10.1109/ACC.2003.1243471}}.

\bibitem{Kim2007}
D.~J. KIM, V.~T. NGUYEN,
  \href{http://www.ijat.net/journal/view.php?number=396}{Reduction of high
  frequency excitations in a cam profile by using modified smoothing spline
  curves}, Int J Automot Technol 8~(1) (2007) 59--66.
\newline\urlprefix\url{http://www.ijat.net/journal/view.php?number=396}

\bibitem{LEE2000}
I.-K. Lee, Curve reconstruction from unorganized points, Computer Aided
  Geometric Design 17~(2) (2000) 161--177.
\newblock \href {https://doi.org/https://doi.org/10.1016/S0167-8396(99)00044-8}
  {\path{doi:https://doi.org/10.1016/S0167-8396(99)00044-8}}.

\bibitem{Schindler2011}
A.~Schindler, G.~Maier, S.~Pangerl, Exploiting arc splines for digital maps,
  in: 2011 14th International IEEE Conference on Intelligent Transportation
  Systems (ITSC), 2011, pp. 1--6.
\newblock \href {https://doi.org/10.1109/ITSC.2011.6082800}
  {\path{doi:10.1109/ITSC.2011.6082800}}.

\bibitem{Schindler2012}
A.~Schindler, G.~Maier, F.~Janda, Generation of high precision digital maps
  using circular arc splines, in: 2012 IEEE Intelligent Vehicles Symposium,
  2012, pp. 246--251.
\newblock \href {https://doi.org/10.1109/IVS.2012.6232124}
  {\path{doi:10.1109/IVS.2012.6232124}}.

\bibitem{MAIER2014}
G.~Maier, Optimal arc spline approximation, Computer Aided Geometric Design
  31~(5) (2014) 211--226.
\newblock \href {https://doi.org/https://doi.org/10.1016/j.cagd.2014.02.011}
  {\path{doi:https://doi.org/10.1016/j.cagd.2014.02.011}}.

\bibitem{Klass1983}
R.~Klass, An offset spline approximation for plane cubic splines,
  Computer-Aided Design 15~(5) (1983) 297--299.
\newblock \href {https://doi.org/https://doi.org/10.1016/0010-4485(83)90019-2}
  {\path{doi:https://doi.org/10.1016/0010-4485(83)90019-2}}.

\bibitem{PIEGL1987}
L.~Piegl, W.~Tiller, Curve and surface constructions using rational b-splines,
  Computer-Aided Design 19~(9) (1987) 485--498.
\newblock \href {https://doi.org/https://doi.org/10.1016/0010-4485(87)90234-X}
  {\path{doi:https://doi.org/10.1016/0010-4485(87)90234-X}}.

\bibitem{HOSCHEK1992}
J.~Hoschek, Circular splines, Computer-Aided Design 24~(11) (1992) 611--618.
\newblock \href {https://doi.org/https://doi.org/10.1016/0010-4485(92)90072-I}
  {\path{doi:https://doi.org/10.1016/0010-4485(92)90072-I}}.

\bibitem{MEEK1992}
D.~Meek, D.~Walton, Approximation of discrete data by g1 arc splines,
  Computer-Aided Design 24~(6) (1992) 301--306.
\newblock \href {https://doi.org/https://doi.org/10.1016/0010-4485(92)90047-E}
  {\path{doi:https://doi.org/10.1016/0010-4485(92)90047-E}}.

\bibitem{YANG1996}
S.-N. Yang, W.-C. Du, Numerical methods for approximating digitized curves by
  piecewise circular arcs, Journal of Computational and Applied Mathematics
  66~(1) (1996) 557--569, proceedings of the Sixth International Congress on
  Computational and Applied Mathematics.
\newblock \href {https://doi.org/https://doi.org/10.1016/0377-0427(95)00191-3}
  {\path{doi:https://doi.org/10.1016/0377-0427(95)00191-3}}.

\bibitem{PIEGL2002}
L.~Piegl, W.~Tiller, Data approximation using biarcs, Engineering With
  Computers 18 (2002) 59--65.
\newblock \href {https://doi.org/10.1007/s003660200005}
  {\path{doi:10.1007/s003660200005}}.

\bibitem{Heimlich2008}
M.~HEIMLICH, M.~HELD, Biarc approximation, simplification and smoothing of
  polygonal curves by means of voronoi-based tolerance bands, International
  Journal of Computational Geometry \& Applications 18~(03) (2008) 221--250.
\newblock \href {https://doi.org/https://doi.org/10.1142/S0218195908002593}
  {\path{doi:https://doi.org/10.1142/S0218195908002593}}.

\bibitem{Song2009}
X.~Song, M.~Aigner, F.~Chen, B.~Jüttler, Circular spline fitting using an
  evolution process, Journal of Computational and Applied Mathematics 231~(1)
  (2009) 423--433.
\newblock \href {https://doi.org/https://doi.org/10.1016/j.cam.2009.03.002}
  {\path{doi:https://doi.org/10.1016/j.cam.2009.03.002}}.

\bibitem{Fischler1981}
M.~A. Fischler, R.~C. Bolles,
  \href{https://doi.org/10.1145/358669.358692}{Random sample consensus: A
  paradigm for model fitting with applications to image analysis and automated
  cartography}, Commun. ACM 24~(6) (1981) 381–395.
\newblock \href {https://doi.org/10.1145/358669.358692}
  {\path{doi:10.1145/358669.358692}}.
\newline\urlprefix\url{https://doi.org/10.1145/358669.358692}

\bibitem{Drysdale2008}
R.~Drysdale, G.~Rote, A.~Sturm, Approximation of an open polygonal curve with a
  minimum number of circular arcs and biarcs, Comput. Geom. 41 (2008) 31--47.

\bibitem{Madsen2004}
K.~Madsen, H.~Nielsen, O.~Tingleff, Methods for Non-Linear Least Squares
  Problems (2nd ed.), Technical University of Denmark, 2004.

\bibitem{Davis2004}
T.~A. Davis, J.~R. Gilbert, S.~I. Larimore, E.~G. Ng,
  \href{https://doi.org/10.1145/1024074.1024080}{Algorithm 836: Colamd, a
  column approximate minimum degree ordering algorithm}, ACM Trans. Math.
  Softw. 30~(3) (2004) 377–380.
\newblock \href {https://doi.org/10.1145/1024074.1024080}
  {\path{doi:10.1145/1024074.1024080}}.
\newline\urlprefix\url{https://doi.org/10.1145/1024074.1024080}

\bibitem{OptToolDoc}
{The MathWorks Inc.},
  \href{https://kr.mathworks.com/help/optim/ug/lsqnonlin.html}{{lsqnonlin :
  Nonlinear Least Squares Solver}} (2023).
\newline\urlprefix\url{https://kr.mathworks.com/help/optim/ug/lsqnonlin.html}

\bibitem{Coleman1994}
T.~F. Coleman, Y.~Li, \href{https://doi.org/10.1137/0806023}{An interior trust
  region approach for nonlinear minimization subject to bounds}, SIAM Journal
  on Optimization 6~(2) (1996) 418--445.
\newblock \href {http://arxiv.org/abs/https://doi.org/10.1137/0806023}
  {\path{arXiv:https://doi.org/10.1137/0806023}}, \href
  {https://doi.org/10.1137/0806023} {\path{doi:10.1137/0806023}}.
\newline\urlprefix\url{https://doi.org/10.1137/0806023}

\bibitem{Pearson1900}
K.~Pearson, On the criterion that a given system of deviations from the
  probable in the case of a correlated system of variables is such that it can
  be reasonably supposed to have arisen from random sampling, The London,
  Edinburgh, and Dublin Philosophical Magazine and Journal of Science 50~(302)
  (1900) 157--175.
\newblock \href {https://doi.org/https://doi.org/10.1080/14786440009463897}
  {\path{doi:https://doi.org/10.1080/14786440009463897}}.

\bibitem{Openpilot}
Comma.ai, inc., openpilot, \url{https://github.com/commaai/openpilot} (2018).

\bibitem{Jeon2023}
J.~Jeon, Robust vehicle trajectory reconstruction and parameterized lane map
  generation via multi-modal sensor fusion, Master's thesis, Korea Advanced
  Institute of Science and Technology(KAIST), Daejeon, South Korea (February
  2023).

\bibitem{Lima2015}
P.~F. Lima, M.~Trincavelli, J.~Mårtensson, B.~Wahlberg, Clothoid-based model
  predictive control for autonomous driving, in: 2015 European Control
  Conference (ECC), 2015, pp. 2983--2990.
\newblock \href {https://doi.org/10.1109/ECC.2015.7330991}
  {\path{doi:10.1109/ECC.2015.7330991}}.

\bibitem{Sparseinv}
Y.~Campbell, T.~Davis, Computing the sparse inverse subset: An inverse
  multifrontal approach, Tech. rep., University of Florida (1995).

\bibitem{MultiDelta}
J.~L. Doob, \href{https://doi.org/10.1214/aoms/1177732594}{{The Limiting
  Distributions of Certain Statistics}}, The Annals of Mathematical Statistics
  6~(3) (1935) 160 -- 169.
\newblock \href {https://doi.org/10.1214/aoms/1177732594}
  {\path{doi:10.1214/aoms/1177732594}}.
\newline\urlprefix\url{https://doi.org/10.1214/aoms/1177732594}

\bibitem{Forster2015}
C.~Forster, L.~Carlone, F.~Dellaert, D.~Scaramuzza,
  \href{http://dx.doi.org/10.15607/rss.2015.xi.006}{Imu preintegration on
  manifold for efficient visual-inertial maximum-a-posteriori estimation}, in:
  Robotics: Science and Systems XI, RSS2015, Robotics: Science and Systems
  Foundation, 2015.
\newblock \href {https://doi.org/10.15607/rss.2015.xi.006}
  {\path{doi:10.15607/rss.2015.xi.006}}.
\newline\urlprefix\url{http://dx.doi.org/10.15607/rss.2015.xi.006}

\end{thebibliography}





% \end{thebibliography}
\end{document}